\documentclass[journal]{IEEEtran}

\usepackage{cite}
\usepackage{amsmath,amssymb}
\usepackage{graphicx}
\usepackage{booktabs}
\usepackage{siunitx}
\usepackage{url}
\usepackage{xcolor}
\usepackage{multirow}
\usepackage{array}
\usepackage{enumitem}
\usepackage{cuted}

\title{Polarimetric SAR Model Fitting for Soil Moisture Retrieval: Study of PALSAR-2 data over a Heterogeneous Mine Environment in Finland}

\author{Oleg~Antropov,
Alireza~Hamedianfar,
Matthieu~Molinier,
Ulla~Salmela,
Hanna~Kukkula,
Lauri~Seitsonen,
Pauliina~Liwata-Kentt\"al\"a,
and Maarit~Middleton%

\thanks{This work was supported by the European Union's Horizon Europe programme under the MultiMiner project, Grant Agreement No. 101091374.}
\thanks{O. Antropov, M. Molinier, and L. Seitsonen are with VTT Technical Research Centre of Finland, Espoo, Finland.}
\thanks{A. Hamedianfar, P. Liwata-Kentt\"al\"a, and M. Middleton are with the Geological Survey of Finland, Finland.}
\thanks{U. Salmela and H. Kukkula are with Nordkalk Oy Ab, Finland.}
}

\begin{document}
\maketitle

\begin{abstract}
This paper examines several model based approaches for retrieving surface soil moisture from ALOS-2 PALSAR-2 quad-pol imagery, over a lime stone quarry in southeastern Finland. The study primarily targets physically interpretable semi-empirical modeling approaches, with generic ML modeling used as a benchmark. Along with common polarimetric observables, we  propose a generalization of the SAR time series based TU Wien soil moisture index (SMI) retrievals examined across several representational spaces derived from polarimetric coherency matrix $[T3]$.
This study was conducted over a closed tailing storage facility and a landfill, with a set of 9 repeat pass PALSAR-2 images. The best semi-empirical configuration combining temporal context SMI and current observation PolSAR parameters achieved $R^2=0.67$ and RMSE $=5.65$ volumetric \% units. The strongest $SMI_{[T3]}$ approach with sediment-specific calibration, achieved $R^2=0.66$ and RMSE $=5.67$ vol. \%, which was considerably better than using $SMI_{HH}$ or $SMI_{VV}$. The proposed approach was sensitive to representations: dB-based projection outperformed linear or trace-normalized $[T3]$ representation. Factoring in sediment information dramatically improved retrieval performance compared to using global model fitting. Machine learning results closely approached but not outperformed semi-empirical model based methodologies. Similarly, they highlighted the need for sediment-specific modeling as well as the importance of including time-series/temporal backscatter dynamics during SSM retrieval. Our study demonstrated the utility of physics based  SSM retrieval approaches in the complex multi-sediment mine environment under relatively scarce reference data conditions. 
\end{abstract}

\begin{IEEEkeywords}
synthetic aperture radar, polarimetry, L-band, surface soil moisture, mining site, coherency matrix, volume scattering, semi-empirical model, machine learning, sediment-specific calibration.
\end{IEEEkeywords}

\section{Introduction}

Surface soil moisture (SSM) is a key environmental state variable controlling evapo-transpiration, vegetation moisture stress, freeze-thaw response, dust emissions, and near-surface bearing capacity. In constructed terrains common e.g. at mine sites, near-surface moisture is also relevant for identifying seepage-prone areas, wetness anomalies on embankments and cover structures, success of revegetation, controlling mineral dust emissions of bare surfaces, and trafficability. Because \textit{in situ} soil moisture sensors provide accurate but spatially sparse measurements, remotely sensed SSM products are attractive as a complementary source of spatial information.

Passive microwave missions, such as SMOS and SMAP, provide valuable regional to global soil moisture products, but their kilometer-scale spatial resolution is generally too coarse for mine site monitoring. Synthetic aperture radar (SAR), in contrast, can provide observations at metre- to decametre-scale resolution and is therefore better suited to heterogeneous landscapes and engineered surface materials. L-band SAR is especially attractive because its longer wavelength is sensitive to near-surface dielectric properties and can retain some soil sensitivity under rough or partially vegetated conditions. When full polarimetry is available, additional observables related to surface, double-bounce, and volume scattering can also be derived and used for describing target properties via change of polarimetric state of electromagnetic wave \cite{Hajnsek2003,Hajnsek2009,Jagdhuber2013,Shi2021}.

Generally, SSM retrieval from SAR is challenging because the observed radar signature is not controlled by soil moisture alone. Surface roughness, soil or sediment texture, local incidence angle, vegetation attenuation, volume scattering, double-bounce interactions, and the geometry of scattering elements all affect the measured backscatter and polarimetric covariance or coherency matrix. In connection with this, SAR-based SSM retrieval has generally followed four methodological directions: semi-empirical modeling, change-detection or temporal-normalization methods, forward modeling based on physical scattering models, and data-driven ML approaches.

Semi-empirical methods establish strong relationships between radar backscatter, soil dielectric permittivity, surface roughness, and vegetation descriptors. 
Classical bare-soil retrieval approaches therefore rely on empirical or semi-empirical relationships between radar backscatter, surface roughness, and soil moisture \cite{Ulaby1978,Oh1992,Ulaby1996} and have been widely used because of their relatively simple parameterization and practical readiness for inverting \cite{Oh1992,Dubois1995,Ulaby1996,Shi1997}. Under vegetation, semi-empirical radiative-transfer approximations such as the Water Cloud Model \cite{Attema1978}, introduce canopy attenuation and volume scattering terms related to random or oriented vegetation, but their performance often depends strongly on empirical coefficients and ancillary vegetation descriptors \cite{Attema1978,ElHajj2016,Singh2023,Yadav2020}. Furthermore, vegetation can attenuate soil contribution, add cross-polarized volume scattering, and create dihedral interactions between stems and the ground. Model-based polarimetric decomposition provides a physically motivated route to separate these effects. However, conventional decomposition models often depend on simplified component assumptions and may become under-determined when only a limited number of observations is available \cite{Freeman1998,Yamaguchi2005,vanZyl2011}.

Change-detection approaches, instead of explicitly modeling all scattering contributions, assume that roughness and vegetation effects are sufficiently stable over a given time window and attribute the dominant temporal variation in observed SAR backscatter to changes in soil moisture. This idea underlies the well known dry--wet normalization approaches and is attractive for operational retrieval because it reduces the need for explicit roughness or vegetation parameterization \cite{Wagner1999,Balenzano2011,Balenzano2021,Zhu2022}. However, its validity depends on the availability of a sufficiently representative time series and on the assumption that non-moisture scattering factors do not change too rapidly over time. For this reason,  this approach was most widely implemented only with Sentinel-1 imagery to date.  

Forward-modeling approaches aim at representing the physical scattering mechanisms more explicitly. Model-based PolSAR decomposition can exploit the polarimetric covariance or coherency matrix to separate surface, volume, and double-bounce contributions, thereby improving the physical interpretability of soil moisture retrieval under vegetation \cite{Freeman1998,Hajnsek2009,Jagdhuber2013,He2016,Wang2017}. 
Hajnsek et al. \cite{Hajnsek2003} introduced the X-Bragg surface model to account for surface depolarization beyond the narrow validity range of the classical Bragg approximation. Subsequent studies have then utilized the X-Bragg for SSM retrieval under vegetation cover \cite{Hajnsek2009,Jagdhuber2013}. Jagdhuber \cite{Jagdhuber2016} further proposed an extended Fresnel formulation for depolarizing double-bounce scattering, allowing soil-vegetation dihedral interactions to be represented more realistically. Shi et al. \cite{Shi2021} combined the X-Bragg surface model, an extended double Fresnel model, and the generalized volume scattering model (GVSM) \cite{antropov2011} to retrieve soil moisture from L-band multi-incidence and multitemporal PolSAR observations over agricultural fields.
These developments highlight the value of physically constrained scattering separation, but also illustrate a central difficulty: the number of unknown surface, vegetation, and structural parameters can easily exceed the information content of sparse SAR observations.

In parallel, machine learning (ML) approaches have become increasingly common for SAR-based soil moisture retrieval. Random forests, support vector regression, gradient boosting, neural networks, and hybrid data-fusion models can integrate multi-source predictors and nonlinear relationships without requiring an explicit scattering model. These methods are attractive when large training datasets are available, and they have shown promising results for Sentinel-1, PALSAR-2, UAVSAR, and multi-source SSM mapping \cite{Abowarda2021,Huang2021Agronomy,Bhogapurapu2022,Fan2025}. However, purely data-driven models may be sensitive to training-data representativeness, site-specific correlations, spatial leakage, and changes in surface or vegetation conditions. Recent reviews and benchmarking studies have therefore emphasized the need to assess generalization and to compare ML retrievals against physics based semi-empirical baselines \cite{Karthikeyan2017,Kornelsen2013,Lamichhane2025,Zhu2025TGRS}.

Most model-based L-band PolSAR SSM studies have been developed and validated over agricultural environments, where field boundaries, crop types, vegetation structure, and management conditions are often better defined than in mine environments. In contrast, constructed geomaterials can contain abrupt transitions e.g. from flotation sand and gravel to cover materials such as organic top soils or clay. They may also have crusted surfaces and uneven vegetation cover. These sediments differ not only in dielectric and roughness properties, but also in hydraulic conductivity, water retention, compaction, poor drainage caused surface water pools, and vegetation establishment. As a result, the same radar observable may correspond to different volumetric soil moisture values in different sediment classes. This motivates sediment-specific retrieval and a careful assessment of which physically interpretable SAR observables remain useful in such a complex setting.

The present study addresses this task using nine repeat-pass ALOS-2 PALSAR-2 quad-pol observations acquired in the same imaging geometry over a spatially heterogeneous mine environment in southeastern Finland. The constant geometry is especially advantageous for temporal change detection and dry--wet normalization because acquisition-angle effects are reduced, while the repeat-pass stack provides temporal information on wetting and drying. The available quad-pol data allows to fully exploit potential of various polarimetric-descriptor representations for improving SSM retrieval performance.

To fully assess potential of various mentioned approaches, we explore SSM retrieval via several targeted groups of methodologies introduced above. The first group consists of compact semi-empirical and physically interpretable scalar models, including HH-based temporal soil moisture indices, type polarization ratios, incidence-angle terms, radar vegetation indices and vegetation scatter contribution terms and corrections. The second group extends the well established dry--wet SAR image normalization concept of \textit{Soil Moisture Index} to various polarimetric coherency/covariance matrix descriptor representations. In this group, each repeat-pass observation is projected onto a dry--wet trajectory estimated from the training data. Several representations are compared, including raw linear full-($[T3]$) matrix projection, trace-normalized ($[T3]$) projection, ($[T3]$) diagonal dB projection, $[T3]_{HH/HV/VV,dB}$ space projection, and compact full-polarimetric descriptor-space projection. The third group consists of generic ML regression approaches (ridge, kNN, Random Forests, SVR, XGBoost), included as an empirical baseline.

This hierarchy of modeling approaches allows us to test not only which method performs best, but also which physical representation of the L-band PolSAR response is most useful for SSM retrieval in a spatially highly variable mine environment. In particular, the comparison between SMI-features derived from various coherency matrix representations allows to assess whether the useful wetness signal is primarily expressed in the full complex coherency matrix, in normalized polarimetric shape, or in visible backscatter-amplitude dynamics.

The objective of this paper is to systematically compare several physically motivated retrieval strategies for repeat-pass ALOS-2 PALSAR-2 quad-polarimetric data, propose possible improvements, and compare versus to generic ML baselines. Based on compiled PALSAR-2 quad-pol feature bank matched with \textit{in situ} SSM measurements over five distinct mine sediment classes: open clay, flotation sand, organic cover with short grass, organic cover with thick grass, and open gravel, we: 1) benchmark a set of physically based semi-empirical retrieval approaches; 2) generalize time-series change detection based SSM retrieval approach in various polarimetric descriptor spaces such as raw coherency matrix ($[T3]$), its trace-normalized version and $dB$ power format, and compact descriptor spaces; 3) compare performance gains versus generic ML approaches on a mine site where sparce labelling data is available.

The remainder of this article is organized as follows. Section II describes the study area, PALSAR-2 imagery and reference data. Section III presents the methodology, including
semi-empirical SSM retrieval approaches, PolSAR-time-series based  projection, machine learning baselines, and cross-validation protocol. Section IV reports the retrieval results for the three method families. Section V discusses factors affecting SSM retrieval. Finally, Section VI summarizes the main conclusions and outlines directions for future work.

\section{Study Area, SAR and Reference Data}

The study site is a an active limestone quarry in southeastern Finland located next to the Lappeenranta city ( $61^\circ 11' 35''\mathrm{N},\ 28^\circ 11' 50''\mathrm{E}$). Landscape restoration has been conducted on an abandoned landfill and a closed section of a tailings storage facility. The site acts as a representative testbed for multi-sediment SSM retrieval \cite{antropovigarss2024, hamedianfar2026highresolutionsedimentspecificsurface}. The monitored area contains several contrasting sediment cover materials and vegetation that were grouped into five sediment/land-cover classes: bare clay, flotation sand, organic soil with short grass, organic soil with thick grass, and bare gravel. For the study site, a particular scientific interest lies in its strong spatial heterogeneity and the operational relevance of near-surface moisture for mine site monitoring.

\begin{figure}[htb]
  \centering
\begin{minipage}[b]{0.95\linewidth}
  \centering
 \centerline{\includegraphics[width=1.\textwidth]{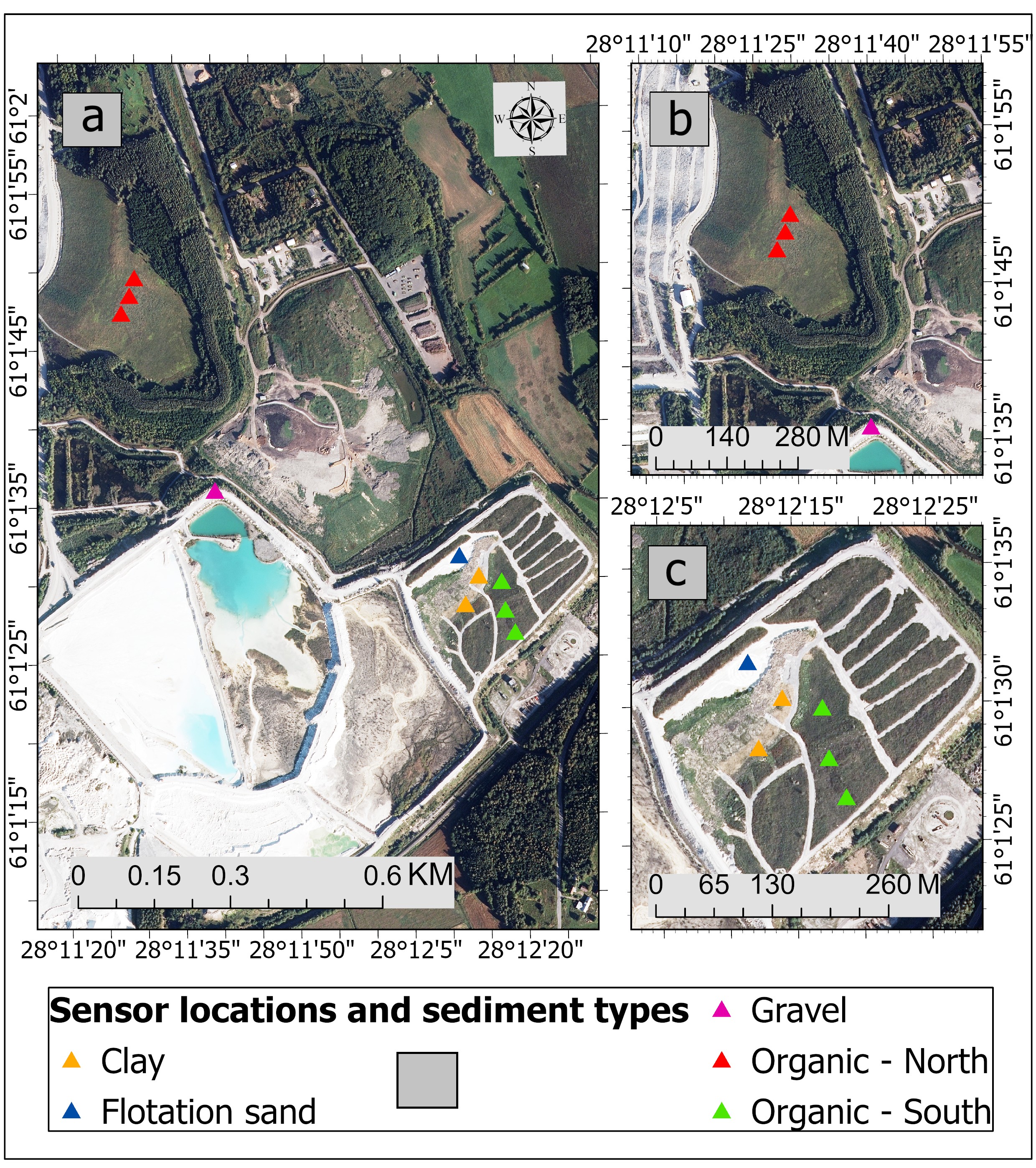}}
  \vspace{0.1cm}
\end{minipage}

\caption{Location of study site within Lappeenranta mine in south-eastern Finland. Locations of IoT soil moisture sensors and sediment types denoted with colour dots (RGB Orthophoto © National Land Survey of Finland): a- overall study area, b - northern site with organic soil and dense grass vegetation, c - multi-sediment site featuring clay dominated and organic soils, as well as gravel and flotation sand areas)}
\label{fig:study_site}

\end{figure}

The SAR dataset consists of nine repeat-pass ALOS-2 PALSAR-2 quad-polarimetric acquisitions acquired in the same imaging geometry between July 27, 2024 and October 4, 2025. The consistent geometry time series are advantageous for temporal normalization and help separating moisture-related temporal change from other scattering effects. Processing started from single look complex (SLC) level 1.1 PALSAR-2 images delivered in CEOS format. The images were polarimetrically calibrated by JAXA \cite{shimada2009}. The pixel spacing of ortho-rectified scenes was set to 10 m. Scenes were multilooked to obtain images with pixel dimensions approximately corresponding to the 10 m cell size. The images were ortho-rectified in the form of Stokes matrix data \cite{rauste2007}. Bi-linear interpolation method was used for resampling in connection with the ortho-rectification. Radiometric normalization of intensity was done using a projected pixel area based approach to minimize topography effects \cite{small2011}. 

Topographic information used in SAR image orthorectification was derived from an airborne laser scanning (ALS) based DEM provided by the National Land Survey of Finland \cite{NLS2025}. The DEM has an pixel size of 2 m, and is referenced to the N2000 height system, and provided in the ETRS89 / TM35FIN coordinate system. Elevation values in the study site ranged from -80.9 to 120.0 m above sea level, with negative values corresponding to the open pit mine. The DEMs were also re-sampled to 10 m pixel spacing using cubic splines.

Ground reference data were obtained from a small IoT-based soil moisture network the sensors installed at the depth of 5 cm below ground surface (shown in Figure 1, see details in \cite{hamedianfar2026highresolutionsedimentspecificsurface}) and matched to the PALSAR-2 observations to form a polarimetric SAR feature bank including non-frozen non-winter images. The final dataset contained 90 PALSAR-2 and reference observation pairs.

\begin{figure*}[htb]
  \centering
\begin{minipage}[b]{0.99\linewidth}
  \centering
 \centerline{\includegraphics[width=1\textwidth]{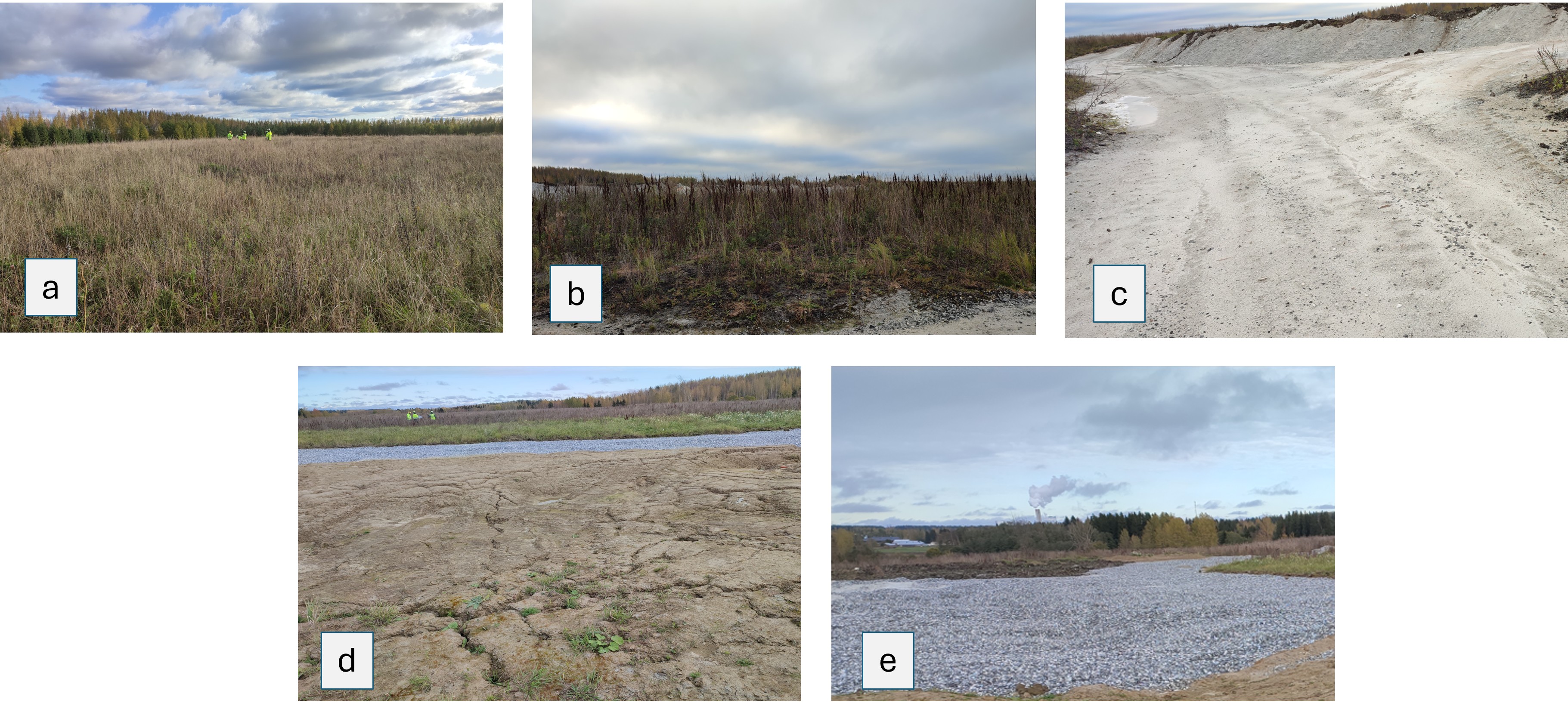}}
  \vspace{0.1cm}
\end{minipage}

\caption{Studied sediment classes with varying vegetation cover: a) Organic soil covered landfill (organic+thick grass), b) organic soil covered tailings storage facility (organic + small grass), c) flotation sand, d) clay, e) gravel.}
\label{fig:sediment_photos}
\end{figure*}

The feature bank contains diagonal powers, polarization ratios, incidence-angle information, commonly used polarimetric descriptors, and full $[T3]$ or $[C3]$-derived terms.

\begin{figure*}[htb]
\centering

\begin{minipage}[b]{0.9\linewidth}
  \centering
 \centerline{\includegraphics[width=1\textwidth]{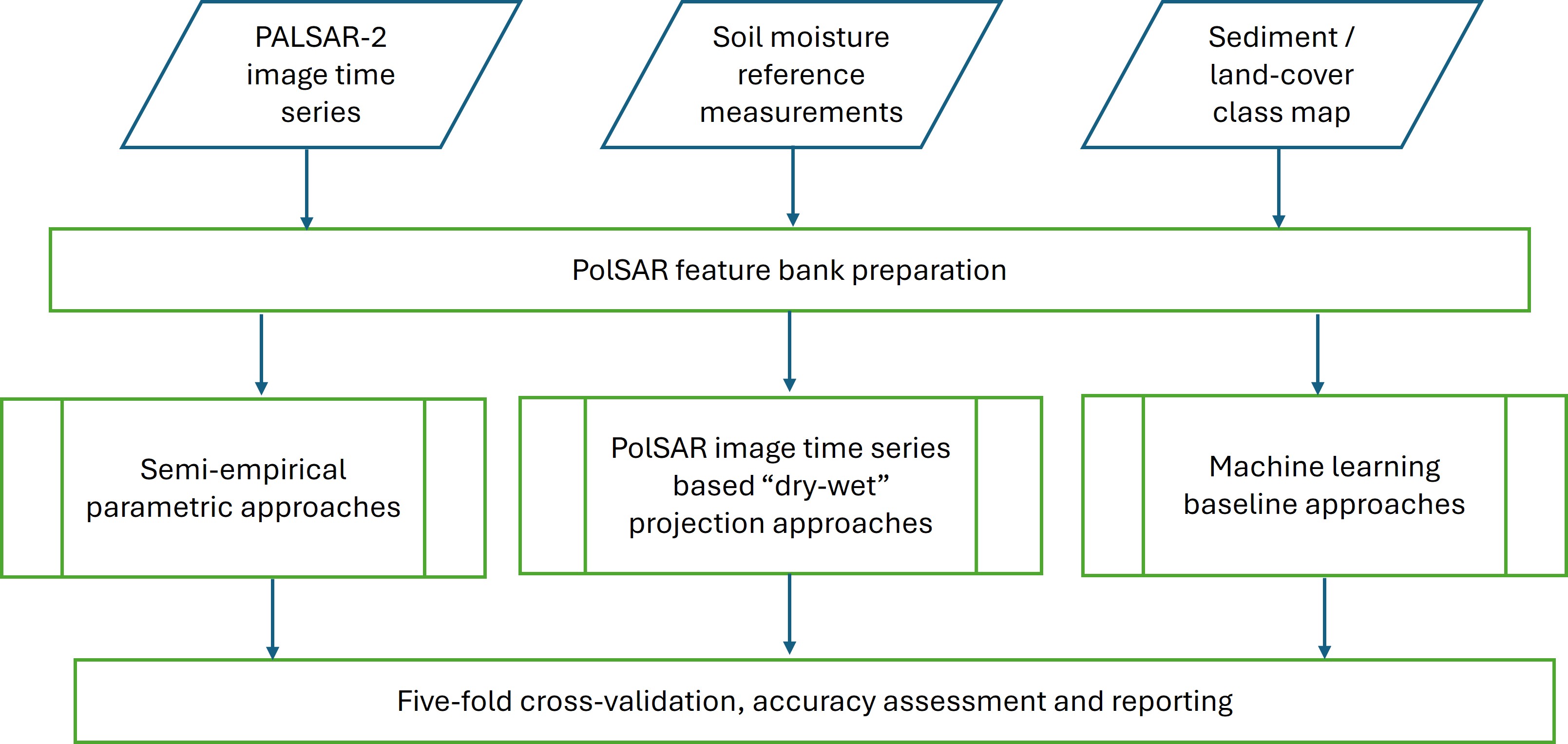}}
  \vspace{0.1cm}
\end{minipage}

\caption{Study logic and hierarchy of evaluated SSM retrieval approaches. Repeat-pass PALSAR-2 observations, in situ soil-moisture measurements, and sediment/cover-class information are combined into a common PolSAR feature bank. Three retrieval families are then evaluated under the same five-fold cross-validation protocol}
\label{fig:studylogic}
\end{figure*}

\section{Methodology}

\subsection{Overall Study logic}

Fig.~\ref{fig:studylogic} summarizes the study logic. The study objective of the methodology is to compare several levels of model complexity under a common validation protocol. The benchmark is organized into three PolSAR based SSM retrieval families. The first family consists of compact semi-empirical and physics-guided scalar retrievals. These methods use physically interpretable polarimetric observables. The second family generalizes the dry--wet normalization concept to multidimensional polarimetric feature spaces, including raw full-($[T3]$), trace-normalized ($[T3]$), dB-scaled ($T3$) diagonal, $HH/HV/VV_{dB}$, and compact full-polarimetric descriptor spaces. The third family consists of generic ML baselines trained on the same feature bank. 
All retrieval methods use the same matched PALSAR-2 and in situ feature bank and are evaluated using the same five-fold cross-validation framework. Whenever calibration is required, including dry/wet reference estimation, affine mapping from index values to SSM, vegetation correction coefficients, or ML model fitting, the parameters are estimated using only the training portion of each fold.

For each acquisition date and sensor location, the quad-polarimetric SAR observation is represented by the complex scattering matrix

\begin{equation}
\mathbf{S} =
\begin{bmatrix}
S_{\mathrm{HH}} & S_{\mathrm{HV}} \\
S_{\mathrm{VH}} & S_{\mathrm{VV}}
\end{bmatrix},
\end{equation}
where monostatic reciprocity is assumed, i.e., ($S_{\mathrm{HV}}=S_{\mathrm{VH}}$).

\noindent The lexicographic covariance matrix is written as
\begin{equation}
\mathrm{[C3]} =
\begin{bmatrix}
\left\langle S_{\mathrm{HH}}S_{\mathrm{HH}}^*\right\rangle &
\left\langle \sqrt{2}S_{\mathrm{HH}}S_{\mathrm{HV}}^*\right\rangle &
\left\langle S_{\mathrm{HH}}S_{\mathrm{VV}}^*\right\rangle \\
\left\langle \sqrt{2}S_{\mathrm{HV}}S_{\mathrm{HH}}^*\right\rangle &
\left\langle 2S_{\mathrm{HV}}S_{\mathrm{HV}}^*\right\rangle &
\left\langle \sqrt{2}S_{\mathrm{HV}}S_{\mathrm{VV}}^*\right\rangle \\
\left\langle S_{\mathrm{VV}}S_{\mathrm{HH}}^*\right\rangle &
\left\langle \sqrt{2}S_{\mathrm{VV}}S_{\mathrm{HV}}^*\right\rangle &
\left\langle S_{\mathrm{VV}}S_{\mathrm{VV}}^*\right\rangle
\end{bmatrix}.
\end{equation}

\newpage
\begin{strip}
\noindent The Pauli coherency matrix is defined as
\begin{equation}
\mathrm{[T3]} =
\frac{1}{2}
\begin{bmatrix}
\left\langle |S_{\mathrm{HH}}+S_{\mathrm{VV}}|^2 \right\rangle &
\left\langle (S_{\mathrm{HH}}+S_{\mathrm{VV}})(S_{\mathrm{HH}}-S_{\mathrm{VV}})^* \right\rangle &
\left\langle 2(S_{\mathrm{HH}}+S_{\mathrm{VV}})S_{\mathrm{HV}}^* \right\rangle \\
\left\langle (S_{\mathrm{HH}}-S_{\mathrm{VV}})(S_{\mathrm{HH}}+S_{\mathrm{VV}})^* \right\rangle &
\left\langle |S_{\mathrm{HH}}-S_{\mathrm{VV}}|^2 \right\rangle &
\left\langle 2(S_{\mathrm{HH}}-S_{\mathrm{VV}})S_{\mathrm{HV}}^* \right\rangle \\
\left\langle 2S_{\mathrm{HV}}(S_{\mathrm{HH}}+S_{\mathrm{VV}})^* \right\rangle &
\left\langle 2S_{\mathrm{HV}}(S_{\mathrm{HH}}-S_{\mathrm{VV}})^* \right\rangle &
\left\langle 4|S_{\mathrm{HV}}|^2 \right\rangle
\end{bmatrix}.
\end{equation}
\end{strip}

\subsection{PALSAR-2 PolSAR Feature Bank}

For each matched PALSAR-2 acquisition and sensor location, a set of polarimetric and auxiliary features was extracted. The core PolSAR features are the diagonal elements of the covariance or coherency representation, denoted here as $C_{11}$, $C_{22}$, and $C_{33}$. They are interpreted as HH, HV, and VV power proxies:
\begin{equation}
    P_{\mathrm{HH}} = C_{11}, \quad
    P_{\mathrm{HV}} = C_{22}, \quad
    P_{\mathrm{VV}} = C_{33}.
\end{equation}

The total span is defined as
\begin{equation}
    \mathrm{Span} = P_{\mathrm{HH}} + P_{\mathrm{HV}} + P_{\mathrm{VV}} .
\end{equation}

Additional features include incidence angle, DEM, selected decomposition fractions, cross-polarized vegetation-sensitive fractions, co-polarized correlation magnitude, and other features listed in Table \ref{tab:polsar_feature_bank}. In our initial experiments, the focus is on compact physically interpretable features that can be used in semi-empirical retrieval models.

\begin{table*}[t]
\centering
\footnotesize
\caption{Polarimetric parameters included in the PALSAR-2 feature bank. The coherency matrix \([T3]\) is represented in the Pauli basis, while the covariance matrix \([C3]\) is represented in the lexicographic basis.}
\label{tab:polsar_feature_bank}
\begin{tabular}{p{3.0cm} p{4.0cm} p{8.9cm}}
\toprule
Feature group & Polarimetric features & Description / interpretation \\
\midrule
Full coherency matrix \(T3\) &
\(T_{11}\), \texttt{\(T_{22}\)}, \texttt{\(T_{33}\)}, \texttt{\(Re\{T_{12}\}\)}, \texttt{\(Im\{T_{12}\}\)}, \texttt{\(Re\{T_{13}\}\)}, \texttt{\(Im\{T_{13}\}\)}, \texttt{\(Re\{T_{23}\}\)}, \texttt{\(Im\{T_{23}\}\)} &
Full polarimetric coherency matrix represented by diagonal powers and real/imaginary parts of the complex off-diagonal terms. These features preserve the complete second-order full-polarimetric information in the Pauli basis. \\
\midrule
Full covariance matrix \(C3\) &
\texttt{\(C_{11}\)}, \texttt{\(C_{22}\)}, \texttt{\(C_{33}\)}, \texttt{\(Re\{C_{12}\}\)}, \texttt{\(Im\{C_{12}\}\)}, \texttt{\(Re\{C_{13}\}\)}, \texttt{\(Im\{C_{13}\}\)}, \texttt{\(Re\{C_{23}\}\)}, \texttt{\(Im\{C_{23}\}\)} &
Full polarimetric covariance matrix in the lexicographic basis. In the present feature bank, \texttt{C11}, \texttt{C22}, and \texttt{C33} correspond to HH-, HV-, and VV-like power channels used for scalar and descriptor-space retrieval experiments. \\
\midrule
Span / total power &
\(Span\) &
Total polarimetric power computed from the trace of \(T_3\) or \(C_3\). These variables describe the overall scattering intensity and can be also used for normalization. \\
\midrule
Normalized \(T3\) diagonal fractions &
\texttt{\(N_{11}=T_{11}/Span\)}, \texttt{\(N_{22}=T_{22}/Span\)}, \texttt{\(N_{33}=T_{33}/Span\)} &
Trace-normalized coherency-matrix diagonal terms. Normalized component fractions interpreted as surface-, double-bounce-, and volume- scattering contributions. These compact physically interpretable descriptors derived from the polarimetric coherency matrix structure \cite{praks2009}. \\
\midrule
Normalized \(C3\) diagonal fractions &
\texttt{\(C_{11}/Span\)}, \texttt{\(C_{22}/Span\)}, \texttt{\(C_{33}/Span\)} &
Trace-normalized covariance-matrix diagonal powers. These describe the relative contributions of the HH-, HV-, and VV- power channels. \\
\midrule
Diagonal power ratios &
\texttt{\(T_{33}/T_{11}\)}, \texttt{\(T_{33}/T_{22}\)} &
Ratios involving the cross-/volume-sensitive \(T_{33}\) component. These features are intended to capture relative depolarized or volume scattering contributions compared with other Pauli-basis components. \\
\midrule
Co-polar coherence and phase 
& \(|\rho_{\mathrm{HHVV}}|\), \(\phi_{\mathrm{HHVV}}\), where \(\rho_{\mathrm{HHVV}} = C_{13}/\sqrt{C_{11}C_{33}}\) and \(\phi_{\mathrm{HHVV}} = \arg(C_{13})\) &
Magnitude and phase of the complex HH--VV correlation. These descriptors are sensitive to the balance between surface, double-bounce, and structural scattering contributions. \\
\midrule
Radar vegetation index (GVSM based) &
\texttt{$RVI_{GVSM}$} &
Radar vegetation polarimetric descriptor used as a compact indicator of depolarization and vegetation/volume scattering strength. As a particular case includes earlier proposed "classical" RVI \cite{kim2009}.  \\

\bottomrule
\end{tabular}
\end{table*}

\subsection{Parametric semi-empirical and physics based approaches}

The first group of SAR based SSM retrieval approaches focuses on physically interpretable scalar/power polarimetric features in the regression modeling.

\subsubsection{HH-based Soil Moisture Index}
 The core baseline is a temporal soil moisture index derived from the HH channel. For each sensor location and training fold, dry and wet reference backscatter levels are estimated from the temporal stack. The HH-based SMI is then defined as

\begin{equation}
    \mathrm{SMI}_{\mathrm{HH}}(t) =
    \frac{\sigma^0_{\mathrm{HH}}(t) - \sigma^0_{\mathrm{HH,dry}}}
    {\sigma^0_{\mathrm{HH,wet}} - \sigma^0_{\mathrm{HH,dry}}}.
\label{eq:hh_smi}
\end{equation}

where ($\sigma^0_{\mathrm{HH,dry}}$) and ($\sigma^0_{\mathrm{HH,wet}}$) are estimated from the training data only. The resulting index is clipped to the interval ([0,1]) and converted to volumetric SSM by linear calibration.
In more details, for scalar and hybrid polarimetric models for SSM retrieval, both global and sediment-specific affine calibrations were considered. In the sediment-specific case, the predicted SSM for sediment class is
\begin{equation}
\widehat{\theta}_{v,c} = a_c z + b_c,
\label{eq:sediment_affine}
\end{equation}
where $\hat{\theta}_{v,c}$ is the predicted volumetric soil moisture for sediment class $c$, and (z) is the scalar index or model-derived proxy, such as, e.g.,  $\mathrm{SMI}_{\mathrm{HH}}$. 

\subsubsection{Oh-Type Polarization-Ratio Model}

To incorporate polarimetric sensitivity to roughness and dielectric changes, an Oh-type ratio formulation is used. The ratios are

\begin{equation}
p =
\frac{
\left\langle |S_{\mathrm{HV}}|^2 \right\rangle
}{
\left\langle |S_{\mathrm{VV}}|^2 \right\rangle
}
=
\frac{P_{\mathrm{HV}}}{P_{\mathrm{VV}}},
\qquad
q =
\frac{
\left\langle |S_{\mathrm{HH}}|^2 \right\rangle
}{
\left\langle |S_{\mathrm{VV}}|^2 \right\rangle
}
=
\frac{P_{\mathrm{HH}}}{P_{\mathrm{VV}}}.
\end{equation}

These ratios were used either directly as explanatory variables or in hybrid formulations together with HH-SMI and incidence angle. The motivation is that co- and cross-polarized ratios partially encode surface roughness, depolarization, and vegetation/volume contributions while remaining compact enough for robust calibration with a limited number of samples.
In simplest case, the fitted model uses logarithmic ratios and local incidence-angle (LIA) $\theta_{LIA}$ dependence:
\begin{equation}
\begin{aligned}
\hat{\theta}_{v,c} &=
    \beta_{0,c}
    + \beta_{1,c}\log(p)
    + \beta_{2,c}\log(q) \\
&\quad
    + \beta_{3,c}\cos(\theta_{LIA})
    + \beta_{4,c}\log(p)\cos(\theta_{LIA}).
\end{aligned}
\end{equation}
Here, $\theta_{LIA}$ is the local incidence angle. Coefficients are estimated separately for each sediment class using ridge regression to stabilize the fitting with limited samples.

\subsubsection{Factoring in Vegetation scattering contribution}

For the grass-covered organic sediments, vegetation attenuation and depolarized volume scattering are expected to modify the soil-sensitive backscatter response. For this purpose, we have tested a role of vegetation descriptor motivated by Generalized volume scattering model (GVSM) \cite{antropov2011} adopted in several PolSAR decomposition approaches, e.g., \cite{Shi2021,mandal2020}, with: 

\begin{equation}
\gamma =
\frac{
\left\langle |S_{\mathrm{HH}}|^2 \right\rangle
}{
\left\langle |S_{\mathrm{VV}}|^2 \right\rangle
}
=
\frac{P_{\mathrm{HH}}}{P_{\mathrm{VV}}}.
\end{equation}

and using the normalized GVSM covariance matrix formulation (\cite{antropov2011}, eq. 9), the volume scattering power can be expressed as

\begin{equation}
f_{v,\mathrm{GVSM}} =
\left\langle |S_{\mathrm{HV}}|^2 \right\rangle
\frac{9(1+\gamma)-2\sqrt{\gamma}}
{3(1+\gamma)-2\sqrt{\gamma}} .
\end{equation}

Then GVSM-based Radar Vegetation Index is defined as

\begin{equation}
\begin{aligned}
\mathrm{RVI}_{\mathrm{GVSM}}
&=
\frac{
2\left\langle |S_{\mathrm{HV}}|^2 \right\rangle
\left[9(1+\gamma)-2\sqrt{\gamma}\right]
}
{
\left[3(1+\gamma)-2\sqrt{\gamma}\right]
\left(
\left\langle |S_{\mathrm{HH}}|^2 \right\rangle
+
\left\langle |S_{\mathrm{VV}}|^2 \right\rangle
+
2\left\langle |S_{\mathrm{HV}}|^2 \right\rangle
\right)
}.
\end{aligned}
\end{equation}

Setting $\gamma=1$ provides the well-known "classical" quad-pol RVI proposed in \cite{kim2009}:

\begin{equation}
\mathrm{RVI}_{\mathrm{GVSM}}\big|_{\gamma=1}
=
\frac{
8\left\langle |S_{\mathrm{HV}}|^2 \right\rangle
}
{
\left\langle |S_{\mathrm{HH}}|^2 \right\rangle
+
\left\langle |S_{\mathrm{VV}}|^2 \right\rangle
+
2\left\langle |S_{\mathrm{HV}}|^2 \right\rangle
}.
\end{equation}

For vegetation-covered organic sediment classes, the volumetric scattering descriptor was included as an additional predictor, whereas for the bare sediment classes clay, flotation sand, and gravel, it was set to zero. Corresponding cover-specific model was

\begin{equation}
\hat{\theta}_{v,c} =
\beta_{0,c}
+ \beta_{1,c}\mathrm{SMI}_{\mathrm{HH}}
+ \beta_{2,c} \log(\mathrm{RVI_{GVSM}}),
\end{equation}
An extended version additionally included Oh-type polarization ratios and local incidence-angle dependence:

\begin{equation}
\begin{aligned}
\hat{\theta}_{v,c}
&=
\beta_{0,c}
+\beta_{1,c}\mathrm{SMI}_{\mathrm{HH}}
+\beta_{2,c}\log(p)
+\beta_{3,c}\log(q) + \\
&\quad
+\beta_{4,c}\cos(\theta)
+\beta_{5,c} \log(\mathrm{RVI_{GVSM}}) .
\end{aligned}
\end{equation}

All coefficients were estimated separately for each sediment class using only the training portion of each relevant k-fold split. 
In addition, an auxiliary diagnostic vegetation-corrected HH-polarized power feature was formed as
\begin{equation}
P_{\mathrm{HH,g}} =
\max\left(P_{\mathrm{HH}} - k_c P_{\mathrm{HV}},\epsilon\right),
\label{eq:decontam_hh}
\end{equation}
where ($k_c$) is a correction coefficient estimated in the training fold and ($\epsilon$) is a small positive constant used to avoid nonphysical negative powers. This experiment tests whether removing a cross-polarized contribution from co-pol power magnitude improves the sensitivity to soil surface scattering contribution. The corresponding featuremay be used within change-detection approaches when calculating vegetation corrected temporal soil moisture index $\mathrm{SMI}_{\mathrm{HH-k*HV}}$.

\subsection{Polarimetric coherency matrix PolSAR-time-series based wet-dry projection}

Here, we generalize the dry--wet normalization concept from a single channel to multidimensional polarimetric feature spaces. Each observation is represented by a vector ($\mathbf{x}_t$), and dry and wet reference vectors are estimated from the training stack. The dry--wet projection coordinate is then
\begin{equation}
p_t =
\frac{
\left(\mathbf{x}_t-\mathbf{x}_{\mathrm{dry}}\right)^T
\left(\mathbf{x}_{\mathrm{wet}}-\mathbf{x}_{\mathrm{dry}}\right)
}{
\left|\mathbf{x}_{\mathrm{wet}}-\mathbf{x}_{\mathrm{dry}}\right|^2
}.
\label{eq:projection_vector}
\end{equation}

\begin{figure}[htb]
  \centering
\begin{minipage}[b]{0.9\linewidth}
  \centering
 \centerline{\includegraphics[width=1.\textwidth]{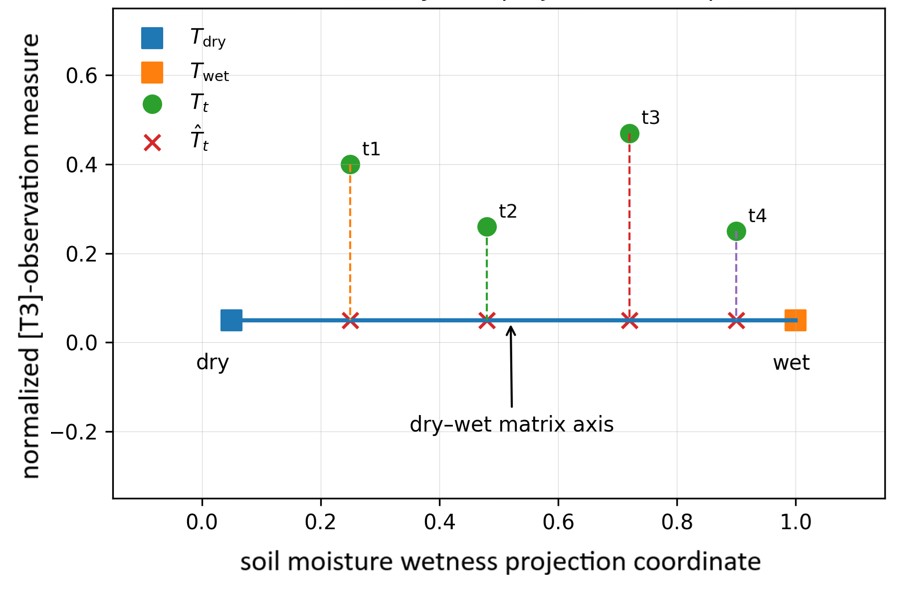}}
  \vspace{0.1cm}
\end{minipage}

\caption{Conceptual illustration of the wet--dry projection approach on polarimetric coherency matrix}
\label{fig:T3concept_figure}
\end{figure}

The coordinate ($p_t$) describes the relative position of polarimetric observation $\mathbf{x}_t$ at the time of the image acquisition $t$ along the estimated dry--wet trajectory. Corresponding SAR-wetness index coordinate is clipped to ([0,1]) and mapped to volumetric SSM using global or sediment-specific affine calibration as in Eq. \eqref{eq:sediment_affine}. A schematic illustration of proposed approach is shown in Figure \ref{fig:T3concept_figure}.

For full matrix-space formulations, the observation is the coherency matrix ($[T3]_t$). The raw full-($[T3]$) projection is computed using the Frobenius inner product:

\begin{equation}
p_{[T3],t} =
\frac{
\left\langle [T3]_t-[T3]_{\mathrm{dry}},
[T3]_{\mathrm{wet}}-[T3]_{\mathrm{dry}}\right\rangle_F
}{
\left|[T3]_{\mathrm{wet}}-[T3]_{\mathrm{dry}}\right|_F^2
},
\label{eq:t3_projection}
\end{equation}
where
\begin{equation}
\langle A,B\rangle_F =
\mathrm{Re}\left(\mathrm{tr}\left(A^{H}B\right)\right).
\end{equation}
This represents direct full-polarimetric extension of the scalar dry--wet concept to full coherency matrix.
To test whether projection should emphasize polarimetric matrix shape rather than total power, a trace-normalized matrix was also used:
\begin{equation}
\widetilde{[T3]}_t =
\frac{[T3]_t}{\mathrm{tr}([T3]_t)}.
\label{eq:trace_norm_t3}
\end{equation}

Dry--wet projection was then applied either to the full trace-normalized matrix or to its normalized diagonal terms. These variants test whether relative polarimetric structure is more informative than absolute amplitude dynamics. As preliminary testing indicated that logarithmic amplitude dynamics were important, dB-scaled descriptor-space projections were also evaluated. The tested projection vectors included

\begin{equation}
\mathbf{x}_{[T3],\mathrm{dB}} =
\left[
T_{11,\mathrm{dB}},
T_{22,\mathrm{dB}},
T_{33,\mathrm{dB}}
\right]^T,
\label{eq:t3diag_db}
\end{equation}
\begin{equation}
\mathbf{x}_{\mathrm{HH/HV/VV},\mathrm{dB}} =
\left[
HH_{\mathrm{dB}},
HV_{\mathrm{dB}},
VV_{\mathrm{dB}}
\right]^T,
\label{eq:hh_hv_vv_db}
\end{equation}
and a compact full-polarimetric descriptor vector combining dB-scaled diagonal powers with selected normalized polarimetric descriptors (Table I).

\subsection{Machine learning baselines}

A set of conventional ML methods was included as empirical reference models. These baselines were trained either globally or in a sediment-specific manner using the same core feature bank as the physics-inspired approaches. Their role here is not to propose a new ML retrieval framework, but to place the physically interpretable models into context and to assess whether flexible empirical regressors provide a clear advantage when applied to the same PALSAR-2 polarimetric descriptors.

The evaluated models included linear regression, ridge regression, $k$ nearest neighbours (kNN) regression, support vector regression (SVR), Random forest regression, and XGBoost. Linear regression was used as the simplest baseline approach, representing the case where soil moisture is assumed to vary linearly with the selected SAR descriptors. Ridge regression extends this model by adding $L_2$ regularization to the regression coefficients, which stabilizes the estimates when correlated polarimetric features are used or when the number of training samples per sediment class is limited \cite{hoerl1970ridge}. The $k$-nearest-neighbours regressor is a non-parametric local interpolation method, where predictions are obtained from the reference soil moisture values of the closest samples in feature space \cite{cover1967nearest}, widely used in remote sensing tasks, e.g. \cite{tomppo2008,antropov2017, khatami2016}. Support vector regression provides a margin-based nonlinear regression baseline and can represent flexible relationships between PolSAR observables and SSM through kernel-based mapping \cite{smola2004tutorial}. Random Forest regression was used as an ensemble tree-based model that captures nonlinear feature interactions by averaging predictions from multiple randomized decision trees \cite{breiman2001random}. XGBoost was included as a gradient-boosted tree ensemble, providing state-of-the-art nonlinear empirical baseline  \cite{chen2016xgboost}.

All ML baselines were evaluated under the same cross-validation protocol as the physics based semi-empirical models. The input feature configurations were varied to assess the contribution of different polarimetric descriptors, advantage of including SAR-time-series derived SMI features, and value of explicit presence of sediment-class labels (global vs sediment-aware modeling). 
The role of ML experiments was to help clarify if generic nonlinear regressors can outperform the earlier examined physically motivated dry--wet and semi-empirical SSM retrieval approaches. Furthermore, ML experiments were done to determine if the main performance gains originate from model flexibility or from physically meaningful auxiliary information such as sediment class and temporal moisture indices.

\subsection{Cross-validation approach and accuracy metrics}

All retrieval families were evaluated under a common k-fold cross-validation framework, with $k=5$. The full set of observations was partitioned into five mutually exclusive test folds. Within each iteration, one fold was used for testing, while the remaining four folds were used for model training. All training-derived method hyperparameters were estimated strictly within the training folds. These included dry and wet reference states, vegetation correction coefficients, feature-scaling parameters, model coefficients, sediment-specific calibrations, as well as hyperparameters of ML algorithms. The held-out fold was then predicted without using its reference soil moisture values for any calibration step.

After completing the five validation iterations, the predictions from the held-out folds were concatenated into a single out-of-fold SSM prediction vector, which was used for assessment of SSM retrieval accuracy. Thus, each observation contributed exactly once to the final accuracy assessment, and the reported metrics were computed from the pooled out-of-fold predictions rather than from fold-wise averages. Let $\theta_{v,i}$ denote the measured volumetric soil moisture for observation $i$, and let $\hat{\theta}_{v,i}$ denote the corresponding out-of-fold prediction. For $n$ valid paired observations, the root-mean-square error was computed as
\begin{equation}
\mathrm{RMSE}
=
\sqrt{
\frac{1}{n}
\sum_{i=1}^{n}
\left(
\theta_{v,i}
-
\hat{\theta}_{v,i}
\right)^2
}.
\end{equation}

The coefficient of determination was computed from the same pooled out-of-fold predictions as
\begin{equation}
R^2
=
1
-
\frac{
\sum_{i=1}^{n}
\left(
\theta_{v,i}
-
\hat{\theta}_{v,i}
\right)^2
}{
\sum_{i=1}^{n}
\left(
\theta_{v,i}
-
\bar{\theta}_{v}
\right)^2
},
\end{equation}
where
\begin{equation}
\bar{\theta}_{v}
=
\frac{1}{n}
\sum_{i=1}^{n}
\theta_{v,i}
\end{equation}
is the mean of the measured soil moisture values over the evaluated observations.

The bias (systematic deviation) was calculated as the mean prediction error:
\begin{equation}
\mathrm{bias}
=
\frac{1}{n}
\sum_{i=1}^{n}
\left(
\hat{\theta}_{v,i}
-
\theta_{v,i}
\right).
\end{equation}
Positive bias therefore indicates overestimation, while negative bias indicates underestimation. RMSE and bias are reported in volumetric percentage (vol.\%) points.

%RMSE and Bias are reported in volumetric percentage points.

\section{Results}

\subsection{Semi-empirical and physics based SSM retrievals}

This group of semi-empirical models already provide a strong benchmark with accuracies approaching the best reported in prior literature. The best-performing configuration was a sensor-normalized HH-SMI model augmented with a vegetation correction for the organic classes, yielding $R^2=0.67$ and RMSE $=5.65$ vol. \%. Several hybrid physics derived models combining HH-SMI, polarization ratios, incidence angle, and vegetation-sensitive terms achieved nearly identical performance, indicating that much of the useful signal is already captured by temporal normalization of the HH response.

\begin{table}[t]
\caption{SSM retrieval performance metrics for parametric semi-empirical and physics-based models}
\label{tab:iva_kfold_parametric_all}
\centering
%\scriptsize
\setlength{\tabcolsep}{2.2pt}
\begin{tabular}{p{0.7\linewidth}ccc}
\hline
Method & RMSE, vol. \% & $R^2$ & Bias \\
\hline
Cover-specific HH-SMI + GRVI (sed.-spec.) & 5.65 & 0.67 & 0.06 \\
\shortstack[l]{Hybrid HH-SMI + Oh ratios + LIA (sed.-spec.)} & 5.81 & 0.65 & -0.35 \\
\shortstack[l]{Extended cover-specific HH-SMI + ratios\\+ LIA + GRVI (sed.-spec.)} & 5.83 & 0.65 & -0.36 \\
Span-SMI (sed.-spec.) & 5.95 & 0.63 & 0.02 \\
Decontaminated HH-SMI (sed.-spec.) & 5.95 & 0.63 & 0.07 \\
HH-SMI (sed.-spec.) & 5.95 & 0.63 & 0.07 \\
Oh ratio model (sed.-spec.) & 6.04 & 0.62 & -0.24 \\
VV-SMI (sed.-spec.) & 6.22 & 0.60 & -0.11 \\
\shortstack[l]{Extended cover-specific HH-SMI + ratios\\+ LIA + GRVI (global)} & 8.00 & 0.33 & -0.09 \\
\shortstack[l]{Hybrid HH-SMI + Oh ratios + LIA (global)} & 8.18 & 0.30 & 0.00 \\
Oh ratio model (global) & 8.66 & 0.22 & -0.04 \\
Cover-specific HH-SMI + GRVI (global) & 9.43 & 0.07 & -0.06 \\
Span-SMI (global) & 9.43 & 0.07 & 0.05 \\
Decontaminated HH-SMI (global) & 9.50 & 0.06 & 0.07 \\
HH-SMI (global) & 9.50 & 0.06 & 0.07 \\
VV-SMI (global) & 9.67 & 0.02 & 0.05 \\
\hline
\end{tabular}
\end{table}

\begin{figure}[htb]
  \centering
\begin{minipage}[b]{0.9\linewidth}
  \centering
 \centerline{\includegraphics[width=1.\textwidth]{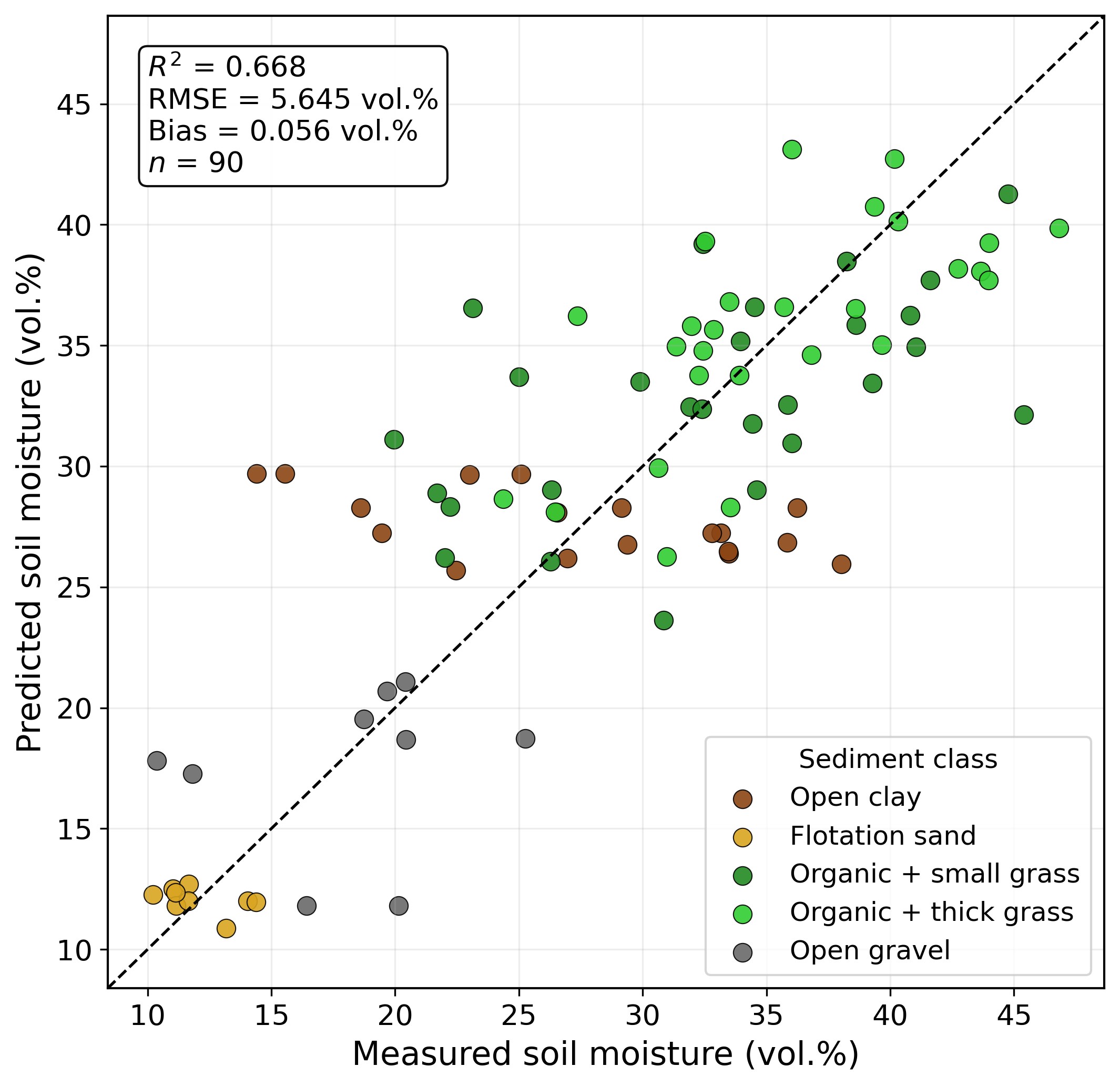}}
  \vspace{0.1cm}
\end{minipage}

\caption{Scatterplot illustrating performance of best approach using semi-empirical models described in Section III-C: Cover-specific HH-SMI + GVSM/RVI model}
\label{fig:SectionA_scatter}
\end{figure}

A representative set of results listed in Table~\ref{tab:iva_kfold_parametric_all} confirms that simple, physically transparent formulations remain highly competitive in the present mine environment. They also indicate that vegetation-specific corrections are useful, but their benefit is relatively modest. 

\subsection{Dry-wet projection of polarimetric coherence matrix in SSM retrievals}

\begin{figure}[htb]
  \centering
\begin{minipage}[b]{0.99\linewidth}
  \centering
 \centerline{\includegraphics[width=1.\textwidth]{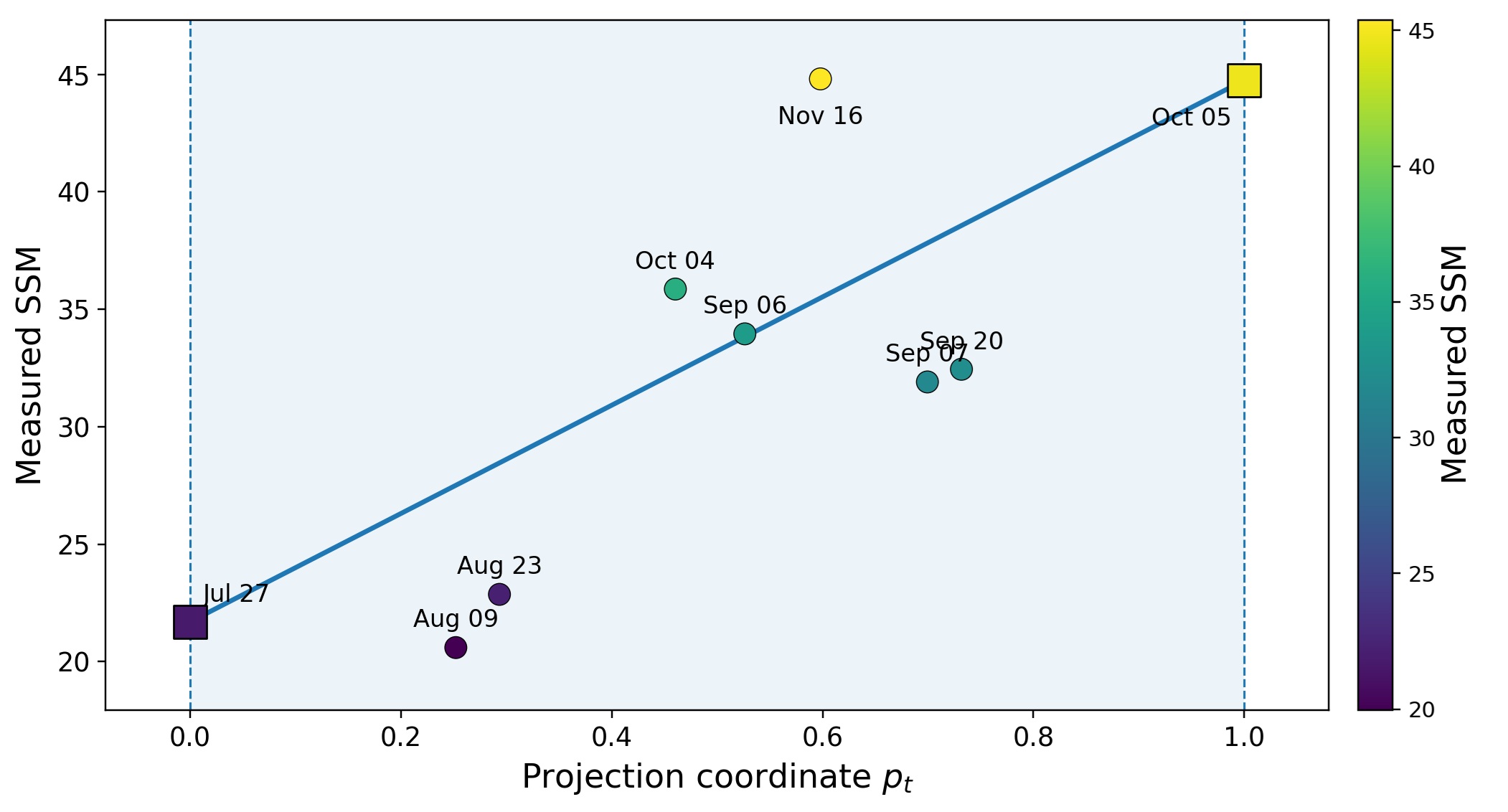}}
  \vspace{0.1cm}
\end{minipage}
\caption{Illustration of the wet--dry matrix projection approach on polarimetric coherency matrix using PolSAR observations from organic class with small/sparse vegetation (Figure 2, b)}
\label{fig:T3projexample}
\end{figure}

Here we show results for various versions of SSM retrieval approaches using dry-wet projection of polarimetric coherency matrix $[T3]$.
We also include an illustration of actual projections in the dry-wet $[T3]$-matrix space, shown in Fig. \ref{fig:T3projexample} for polarimetric observations of class 3 representing organic soil with relatively sparse and smaller grassland vegetation compared to class 4. 

Retrieval results are summarized in Table~\ref{tab:ivb_projection_results}.
Several interesting observations can be made. Firstly, results from dB-space projection were much stronger compared to non-dB space trace normalized projection, indicating that absolute power is more important than polarimetric shapes and orientations, with average difference on order of 1 vol.\%. Secondly, we got surpisingly close performance using polarimetric descriptor containing only HH and HV dB-channels. This is an important finding relevant for example for the NISAR mission, where a majority of stripmap imaging mode datatakes will be collected in exactly this imaging configuration. Adding further polarimetric descriptors, with or without LIA, provided no improvement versus baseline $[T3]$-diagonal dB projection. Major differences were observed for the results of experimental runs with global and sediment-specific calibration, of more than 3 vol. \% difference. This indicates that simple projection is not enough and overall strongly highlights the need for sediment level calibration of retrievals, as mapping from projection coordinate to volumetric SSM is strongly sediment-dependent.

\begin{table}[t]
\caption{SSM retrieval performance metrics for dry--wet coherency matrix projection approaches }
\label{tab:ivb_projection_results}
\centering
%\scriptsize
\setlength{\tabcolsep}{2.2pt}
\begin{tabular}{p{0.6\linewidth}ccc}
\hline
Method & RMSE, vol. \% & $R^2$ & Bias \\
\hline
$[T3]$ diagonal dB projection (sed.-spec.) & 5.67 & 0.66 & -0.13 \\
HH/HV dB projection (sed.-spec.) & 5.67 & 0.66 & -0.19 \\
\shortstack[l]{Compact full-pol descriptor projection (sed.-spec.)} & 5.69 & 0.66 & -0.21 \\
\shortstack[l]{Compact full-pol descriptor\\projection, no LIA (sed.-spec.)} & 5.69 & 0.66 & -0.21 \\
HH/HV/VV dB projection (sed.-spec.) & 5.69 & 0.66 & -0.21 \\
\shortstack[l]{Trace-normalized full-$[T3]$ projection (sed.-spec.)} & 6.43 & 0.57 & -0.47 \\
\shortstack[l]{Trace-normalized $[T3]$ diagonal\\projection (sed.-spec.)} & 6.66 & 0.54 & -0.21 \\
HH/VV dB projection (sed.-spec.) & 6.68 & 0.53 & -0.05 \\
HH/HV/VV dB projection (global) & 9.68 & 0.02 & -0.18 \\
\shortstack[l]{Compact full-pol descriptor \\ projection, no LIA (global)} & 9.69 & 0.02 & -0.18 \\
\shortstack[l]{Compact full-pol descriptor\\projection (global)} & 9.69 & 0.02 & -0.18 \\
HH/HV dB projection (global) & 9.77 & 0.01 & 0.13 \\
$[T3]$ diagonal dB projection (global) & 10.07 & -0.06 & 0.13 \\
HH/VV dB projection (global) & 10.14 & -0.07 & -0.03 \\
\shortstack[l]{Trace-normalized full-$[T3]$projection (global)} & 10.34 & -0.11 & -0.15 \\
\shortstack[l]{Trace-normalized $[T3]$ diagonal projection (global)} & 10.40 & -0.13 & 0.10 \\
\hline
\end{tabular}
\end{table}

\begin{figure}[htb]
  \centering
\begin{minipage}[b]{0.9\linewidth}
  \centering
 \centerline{\includegraphics[width=1.\textwidth]{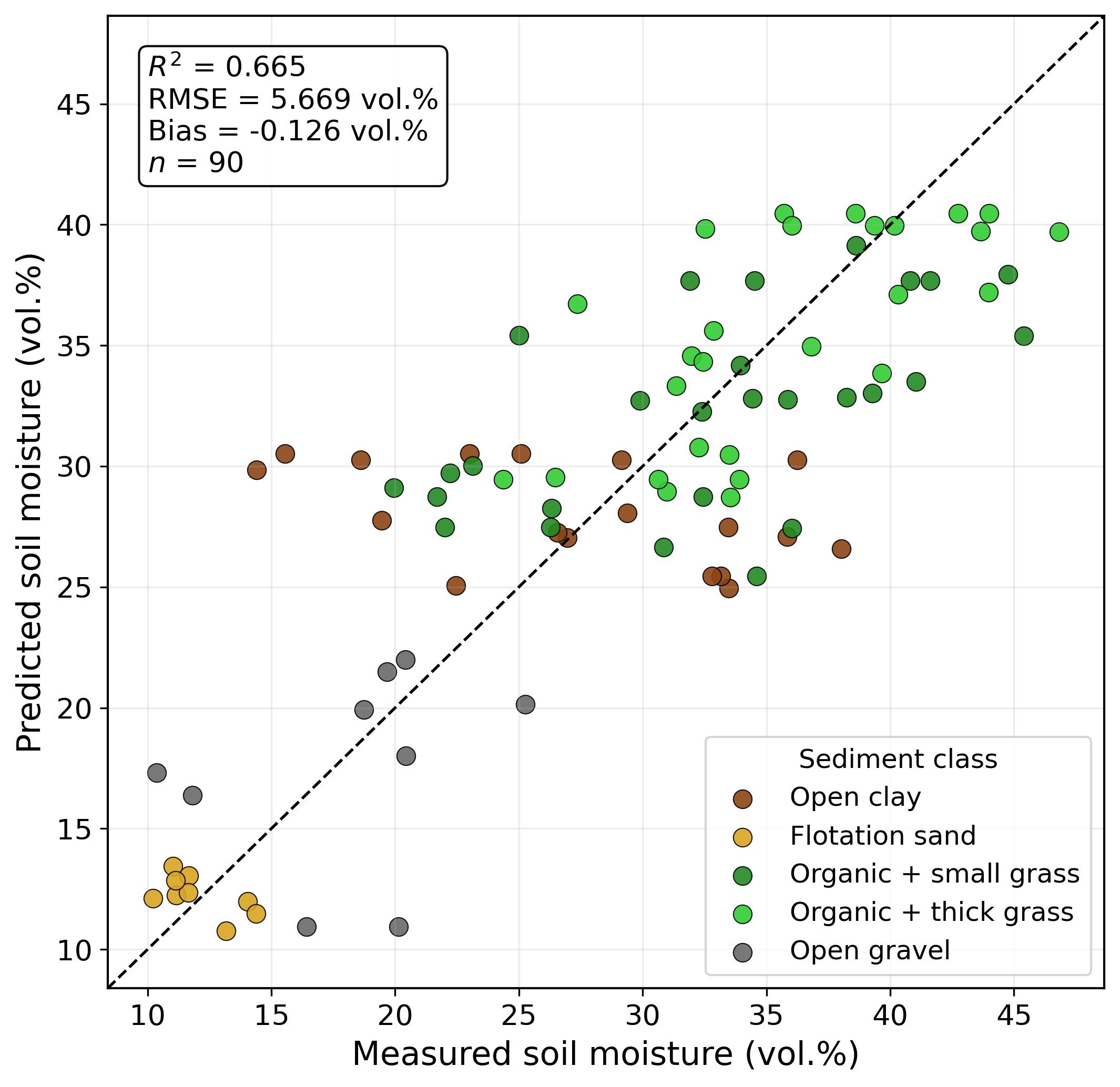}}
  \vspace{0.1cm}
\end{minipage}

\caption{Scatterplot illustrating performance of best approach using temporal projection approaches described in Section III-D using T3 diagonal elements in dB-scale}
\label{fig:SecD_scatter}
\end{figure}

\subsection{Machine learning methods in L-band PolSAR based SSM retrieval}

Here we show results for various versions of ML methods with and without the temporal features, as well as with and without the sediment labeling. Sediment information was supplied as a one-hot encoder predictor rather than class integer numbers. The same matching k-folds were used as in earlier experiments to enable direct comparison of SSM retrieval performance. 

Table~\ref{tab:ml_no_smi_global_sediment_onehot_full} reports single-date ML baselines, while Table V reports temporal ML baselines with fold-wise SMI features. The former uses in principle single-date approaches, whereas the latter focuses on temporal ML experiments where SMI features are computed inside each training fold only and then applied to the corresponding test fold. This avoids data leakage and uses the same fold structure as in previous experiments.

Two predefined single-image feature groups were used for the ML baselines. The first group, denoted as the compact PolSAR feature set, included the backscatter intensities ($\sigma^0_{HH,\mathrm{dB}}$), ($\sigma^0_{HV,\mathrm{dB}}$), and ($\sigma^0_{VV,\mathrm{dB}}$), the logarithmic polarization ratios ($\log(p)$) and ($\log(q)$), where ($p=P_{HV}/P_{VV}$) and ($q=P_{HH}/P_{VV}$), the GVSM/RVI, the normalized volume/depolarized contribution ($T_{33}/\mathrm{Span}$), the magnitude of the HH--VV correlation proxy ($|\rho_{HHVV}|$), and the local incidence angle. The second group, denoted as the extended PolSAR compact feature set, extended the compact feature set with additional descriptors derived from the polarimetric coherency matrix $[T3]$. These included the diagonal terms ($T_{11,\mathrm{dB}}$), ($T_{22,\mathrm{dB}}$), and ($T_{33,\mathrm{dB}}$), the trace-normalized diagonal terms ($N_{11}$), ($N_{22}$), and ($N_{33}$), and the cross-polarized power fraction ($f_{HV}=P_{HV}/{Span})$).

As anticipated, SMI features improve the ML models over the single-date ML runs. Surprisingly, more advanced methods appeared less adequate for regression task, which can be explained by the low number of observations. Thus, there is a limited potential for optimizing the most complex ML approaches. Whereas use of ridge or linear regression with selected polarimetric variables may yield expressions similar to semi-empirical parametric models descried in Section IV-C. 
Global fitting of models without sediment information provided generally weaker SSM retrievals, while adding sediment class as one-hot predictors substantially improves performance. Then, within the sediment-specific setup, adding SMI further improves the best model from R² = 0.616, RMSE = 6.072 to  R² = 0.655, RMSE = 5.748 in the best case and by approximately 1 vol. \% unit on average.

Results presented in Table~\ref{tab:ml_no_smi_global_sediment_onehot_full} support the previously made observation that adding sediment information strongly improves the prediction performance. Expanding the initial PoLSAR feature set brings only a limited improvement because of high inter-correlation between the polarimetric features.

An interesting result is the strong performance of basic regression models such as the ridge regression. The primary explanation for this is likely the small feature set, which can cause a tendency to overfitting within training folds when the cross-validation k-fold approach is used. The best ML results which was based only on single-image descriptors, were obtained by ridge regression with a sediment class included as one-hot encoded categorical predictors. This indicates that, for the available sample size, much of the predictable variation is captured by the sediment-dependent offsets and the linear responses of the SAR descriptors. Benefiting from earlier explored semi-empirical physics inspired approaches, ML results indicate that when reference dataset is very limited, using simpler models can be more useful. It is important to keep in mind that the extended PolSAR descriptor set also contains correlated variables, for which ridge regularization is more robust than the nonlinear models that may be prone to overfitting fold-specific patterns.

Nonparametric regression approaches, commonly praised in literature, including SVR, Random Forest, and Gradient boosting, were not necessarily the best in sediment-specific modeling, probably because of the number of training samples is limited relative to the dimensionality and heterogeneity of the feature space. However, when the sediment information is not provided, advanced ML approaches such as XGBoost and Random Forest being the more competitive than the linear regression or kNN methods, indicating that sediment characterization is indirectly captured by these more flexible models.

\begin{table*}[t]
\caption{ML based SSM retrieval performance with sediment class info but without temporal context of SMI features.}
\label{tab:ml_no_smi_global_sediment_onehot_full}
\centering
\scriptsize
\setlength{\tabcolsep}{2.4pt}
\begin{tabular}{lllcccc}
\hline
Calibration & Method & Feature set & Feat. num. & $R^2$ & RMSE, vol. \% & Bias \\
\hline
Global + sediment one-hot & Ridge regression & Extended PolSAR feature set  + sediment & 21 & 0.616 & 6.072 & -0.180 \\
Global + sediment one-hot & Gradient boosting & Compact PolSAR feature set + sediment & 15 & 0.607 & 6.135 & -0.214 \\
Global + sediment one-hot & Ridge regression & Compact PolSAR feature set + sediment & 15 & 0.605 & 6.151 & -0.194 \\
Global + sediment one-hot & SVR & Compact PolSAR feature set + sediment & 15 & 0.594 & 6.239 & 0.082 \\
Global + sediment one-hot & Linear regression & Compact PolSAR feature set + sediment & 15 & 0.554 & 6.540 & -0.353 \\
Global + sediment one-hot & Gradient boosting & Extended PolSAR feature set + sediment & 21 & 0.545 & 6.606 & -0.471 \\
Global + sediment one-hot & Linear regression & Extended PolSAR feature set + sediment & 21 & 0.544 & 6.611 & -0.293 \\
Global + sediment one-hot & SVR & Extended PolSAR feature set + sediment & 21 & 0.533 & 6.689 & -0.190 \\
Global & Gradient boosting & Compact PolSAR feature set & 10 & 0.501 & 6.917 & -0.147 \\
Global + sediment one-hot & kNN & Compact PolSAR feature set + sediment & 15 & 0.478 & 7.073 & -0.547 \\
Global + sediment one-hot & kNN & Extended PolSAR feature set + sediment & 21 & 0.438 & 7.341 & -0.518 \\
Global + sediment one-hot & Random forest & Extended PolSAR feature set + sediment & 21 & 0.434 & 7.364 & -0.100 \\
Global + sediment one-hot & Random forest & Compact PolSAR feature set + sediment & 15 & 0.410 & 7.521 & -0.222 \\
Global & Gradient boosting & Extended PolSAR feature set & 16 & 0.388 & 7.658 & -0.416 \\
Global & Random forest & Extended PolSAR feature set & 16 & 0.306 & 8.159 & -0.141 \\
Global & Ridge regression & Extended PolSAR feature set & 16 & 0.300 & 8.193 & -0.055 \\
Global & Ridge regression & Compact PolSAR feature set & 10 & 0.269 & 8.373 & 0.066 \\
Global & Random forest & Compact PolSAR feature set & 10 & 0.255 & 8.454 & -0.361 \\
Global & SVR & Compact PolSAR feature set & 10 & 0.225 & 8.618 & 0.646 \\
Global & SVR & Extended PolSAR feature set & 16 & 0.221 & 8.644 & 0.651 \\
Global & Linear regression & Extended PolSAR feature set & 16 & 0.210 & 8.703 & -0.121 \\
Global & Linear regression & Compact PolSAR feature set & 10 & 0.179 & 8.874 & -0.087 \\
Global & kNN & Extended PolSAR feature set & 16 & 0.125 & 9.161 & -1.115 \\
Global & kNN & Compact PolSAR feature set & 10 & 0.032 & 9.635 & -1.668 \\
\hline
\end{tabular}
\end{table*}

\begin{table*}[htb]
\caption{ML based SSM retrieval performance metrics with sediment class info and SMI features.}
\label{tab:ml_smi_global_sediment_onehot_full}
\centering
\scriptsize
\setlength{\tabcolsep}{2.4pt}
\begin{tabular}{lllcccc}
\hline
Calibration & Method & Feature set & Feat. num. & $R^2$ & RMSE, vol. \% & Bias \\
\hline
Global + sediment one-hot & Ridge regression & Extended PolSAR feature set + SMI + sediment & 26 & 0.655 & 5.748 & -0.358 \\
Global + sediment one-hot & Ridge regression & Compact PolSAR feature set + SMI + sediment & 20 & 0.632 & 5.941 & -0.358 \\
Global + sediment one-hot & Linear regression & Compact PolSAR feature set + SMI + sediment & 20 & 0.625 & 5.997 & -0.333 \\
Global + sediment one-hot & Ridge regression & SMI + sediment & 10 & 0.617 & 6.059 & -0.369 \\
Global + sediment one-hot & Gradient boosting & Compact PolSAR feature set + SMI + sediment & 20 & 0.613 & 6.095 & -0.431 \\
Global + sediment one-hot & Linear regression & SMI + sediment & 10 & 0.612 & 6.102 & -0.369 \\
Global + sediment one-hot & Gradient boosting & SMI + sediment & 10 & 0.601 & 6.182 & -0.282 \\
Global + sediment one-hot & kNN & Extended PolSAR feature set + SMI + sediment & 26 & 0.598 & 6.206 & -0.123 \\
Global + sediment one-hot & Linear regression & Extended PolSAR feature set + SMI + sediment & 26 & 0.594 & 6.243 & -0.543 \\
Global + sediment one-hot & SVR & SMI + sediment & 10 & 0.592 & 6.255 & -0.275 \\
Global + sediment one-hot & Gradient boosting & Extended PolSAR feature set + SMI + sediment & 26 & 0.591 & 6.263 & -0.428 \\
Global + sediment one-hot & SVR & Compact PolSAR feature set + SMI + sediment & 20 & 0.584 & 6.314 & -0.424 \\
Global + sediment one-hot & SVR & Extended PolSAR feature set + SMI + sediment & 26 & 0.583 & 6.321 & -0.245 \\
Global + sediment one-hot & kNN & Compact PolSAR feature set + SMI + sediment & 20 & 0.580 & 6.347 & -0.018 \\
Global + sediment one-hot & Random forest & SMI + sediment & 10 & 0.547 & 6.588 & -0.200 \\
Global + sediment one-hot & kNN & SMI + sediment & 10 & 0.541 & 6.637 & -0.341 \\
Global + sediment one-hot & Random forest & Compact PolSAR feature set + SMI + sediment & 20 & 0.503 & 6.907 & -0.240 \\
Global & Gradient boosting & Compact PolSAR feature set + SMI & 15 & 0.482 & 7.046 & -0.300 \\
Global + sediment one-hot & Random forest & Extended PolSAR feature set + SMI + sediment & 26 & 0.437 & 7.347 & -0.395 \\
Global & Gradient boosting & Extended PolSAR feature set + SMI & 21 & 0.422 & 7.445 & -0.509 \\
Global & Ridge regression & Extended PolSAR feature set + SMI & 21 & 0.383 & 7.689 & 0.053 \\
Global & Linear regression & Extended PolSAR feature set + SMI & 21 & 0.321 & 8.070 & -0.206 \\
Global & Ridge regression & Compact PolSAR feature set + SMI & 15 & 0.313 & 8.116 & -0.031 \\
Global & Random forest & Extended PolSAR feature set + SMI & 21 & 0.307 & 8.150 & -0.385 \\
Global & Linear regression & Compact PolSAR feature set + SMI & 15 & 0.288 & 8.262 & -0.040 \\
Global & Random forest & Compact PolSAR feature set + SMI & 15 & 0.275 & 8.338 & -0.394 \\
Global & SVR & Extended PolSAR feature set + SMI & 21 & 0.244 & 8.513 & 0.583 \\
Global & SVR & Compact PolSAR feature set + SMI & 15 & 0.224 & 8.625 & 0.725 \\
Global & kNN & Compact PolSAR feature set + SMI & 15 & 0.137 & 9.096 & -1.139 \\
Global & kNN & Extended PolSAR feature set + SMI & 21 & 0.096 & 9.311 & -1.223 \\
Global & Random forest & SMI & 5 & 0.095 & 9.317 & -0.306 \\
Global & Ridge regression & SMI & 5 & 0.091 & 9.336 & -0.190 \\
Global & Linear regression & SMI & 5 & 0.082 & 9.381 & -0.162 \\
Global & SVR & SMI & 5 & 0.068 & 9.451 & 1.771 \\
Global & kNN & SMI & 5 & 0.002 & 9.783 & 0.192 \\
Global & Gradient boosting & SMI & 5 & -0.041 & 9.992 & -0.160 \\
\hline
\end{tabular}
\end{table*}

\section{Discussion}

\subsection{Sediment heterogeneity as a dominant retrieval factor} 

A major distinction from agricultural studies is the strong role of sediment class. 
Across all evaluated approaches tested for SSM retrieval, sediment-specific calibration substantially improved the retrieval accuracy compared to the globally fitted models. This indicates that the mapping of volumetric SSM from L-band PolSAR observables is strongly sediment dependent. 
This behavior is physically expected for the studied environment. The five monitored sediment cover classes differ not only in their typical moisture ranges, but also in grain-size distribution, compaction, surface roughness, drainage capacity, organic content, and vegetation cover. Consequently, the same measured backscatter levels or polarimetric descriptor values may correspond to different volumetric SSM values depending on the sediment class. For example, bare flotation sand, clay and gravel, and organic sediment can have different porosity, water-retention behavior, and surface structural response during wetting and drying. These differences affect both the dielectric response and the scattering geometry observed by L-band SAR. 
The strong gapin SSM retrieval performance between sediment-specific and global calibrations suggests that a single site-wide retrieval function is not adequate for this type of engineered heterogeneous terrain. In practical terms, the retrieval problem is better understood as a set of sediment- and vegetation cover type-specific inverse problems, rather than as one universal relationship between SAR observables and SSM.  

The role of sediment information was also evident in the ML experiments. Adding sediment class as one-hot encoded predictor substantially improved the performance of the empirical models, whereas models trained without sediment information performed considerably worse ( with RMSE deteriorating appr. 2-3 vol. \% units). This indicates that ML flexibility, especially under scarce reference data like in this study,  does not compensate for missing information about the underlying material class. Instead, sediment class provides physically meaningful contextual information that helps constrain the retrieval problem.

\subsection{Model-based versus machine learning approaches}

The ML baselines provided an empirical reference against which the physics based models could be evaluated. 
Overall, the best ML configurations approached the accuracy of the semi-empirical and time series SMI (dry--wet projection) based methods, but did not outperform them. This is an important result because it suggests that, for the present dataset, physically motivated compact formulations already capture much of the retrievable SSM information. 

Amongst the ML approaches, regularized linear models performed particularly well when sediment information and temporal SMI-type features were included. This can be explained by three factors. First, the number of matched reference observations is limited relative to the dimensionality and heterogeneity of the feature space. Second, many polarimetric descriptors are physically related and therefore correlated. Third, sediment-specific offsets and approximately linear dependencies explain a substantial fraction of the predictable variation. Under these conditions, ridge regression can provide a robust compromise between flexibility and regularization. More flexible nonlinear models, such as random forest, support vector regression, and gradient boosting, were not consistently superior in the sediment-specific experiments. They were however, as expected, better explaining variability of signatures in global fitting. 
This does not imply that such models are generally unsuitable for SAR-based SSM retrieval. Rather, it reflects the constraints of the present dataset: limited sample size, multiple sediment classes, and a relatively short PolSAR time series, paired with chosen k-fold cross-validation approach. In this setting, nonlinear models may fit fold-specific patterns that do not generalize well to held-out observations. In this regards, sometimes simpler models assuming linear dependencies can be better, as is often the case with SAR based modeling where even the presence of speckle can play an important role. Our results advocate the use of physics motivated and regularized models as robust baselines when reference data are sparse. 

Our study also highlights that the major performance gains are not attributed to increased model complexity but rather to physically meaningful contextual information. Sediment class labels and temporal soil moisture indices improved retrieval performance more clearly than simply increasing the flexibility of semi-empirical or machine learning regression model. This finding is consistent with the broader need to combine data-driven methods with physical understanding when applying SAR-based SSM retrieval in complex environments.

\subsection{Role of PolSAR-time-series based Soil moisture Index (dry--wet projection}

The dry--wet projection experiments extend the conventional temporal normalization concept from a single backscatter channel to multidimensional polarimetric descriptor spaces.

The results show that this generalization of TU Wien approach is feasible, but that the chosen representation strongly affects retrieval performance. As such, predictions provided by best [T3]-matrix combination were better than those of HH-SMI approach. 

In particular, dB-scaled power and descriptor-space projections clearly outperformed raw linear matrix-space and trace-normalized matrix-shape formulations. This result suggests that in the present repeat-pass L-band dataset, the most useful moisture-related signal is expressed primarily through logarithmic amplitude dynamics rather than through normalized polarimetric matrix shape alone. The trace-normalized formulations remove much of the total power information and retain mainly relative scattering-mechanism proportions. Their weaker performance indicates that these relative proportions alone are insufficient to describe the wetting and drying trajectory in this environment. In contrast, absolute or logarithmic power changes remain essential for SSM retrieval. Cross-evaluation of various \(T_3\) derived descriptor-space projections indicates that much of the useful temporal wetness signal can be captured by a relatively small set of amplitude polarimetric channels. Adding auxiliary polarimetric descriptors has not readily lead to better SSM retrieval performance. 
It is important to keep in mind also the limited sample size of PolSAR observations, and the fact that many polarimetric descriptors are in general very correlated. It also suggests that, for operational applications, compact dual- or compact-pol-like representations may retain much of the practically useful information, provided that sediment-specific calibration or other contextual constraints are available. 

The dry--wet projection framework is nevertheless useful because it provides an interpretable way to represent multitemporal PolSAR observations as movement along a physically motivated wetness axis. The projection coordinate can be inspected directly, compared between descriptor spaces, and related to measured SSM. This makes the approach more transparent than purely empirical multivariate regression, especially when the objective is not only to maximize accuracy but also to understand which SAR response components carry the moisture-sensitive signal.

\subsection{Role of vegetation scattering correction}

Effects of vegetation scattering  were explicitly considered here because part of the study area is covered by herbaceous vegetation where the radar response is not expected to originate from the soil surface alone. In such areas, vegetation can attenuate the soil contribution, introduce depolarized volume scattering, and modify the relative balance between co- and cross-polarized channels. The semi-empirical experiments therefore tested both vegetation-specific index corrections and a simple cross-polarization-based decontamination of the nominally soil-sensitive HH channel.

The most successful vegetation-specific semi-empirical model was the cover-specific HH-SMI model augmented with the GVSM-inspired radar vegetation descriptor. This model achieved the best performance among the semi-empirical approaches, indicating that the vegetation-sensitive descriptor provides useful additional information when applied to the organic and grass-covered classes. However, the improvement over the uncorrected HH-SMI baseline was relatively modest. This suggests that temporal normalization of HH backscatter already captures much of the dominant wetness-related signal, while the vegetation descriptor mainly helps to account for residual class-dependent vegetation or depolarization effects.

The results also show that the role of the vegetation descriptor is not simply to remove vegetation scattering in a strict physical sense. The RVI-type terms appear more useful as compact empirical indicators of depolarization, canopy/cover structure, and volume-like scattering contributions. These descriptors can help condition the retrieval in vegetated or organic classes, but they do not isolate scattering from soil-surface. This is expected in a heterogeneous mine environment, where cross-polarized terms may also include contributions from small-scale topography. Likewise, auxiliary HH decontamination experiment provided similar results. Subtracting a scaled HV contribution from the HH power targeted reduction of vegetation-related contribution of the co-polarized soil-sensitive channel. However, this approach did not improve the retrieval compared with the best HH-SMI and cover-specific GVSM/RVI-based approaches. A likely reason is that the HV contribution incapsulates information related not only to vegetation volume scattering, but also to soil roughness, sediment structure, and moisture-dependent depolarization. Removing it through a simple linear subtraction can therefore discard useful information. General complexity of separating vegetation induced volume scattering from other scatterimg mechanism contributions is well known and widely explored, in general case requiring multi-incidence angle PolSAR obervations \cite{Shi2021, Hajnsek2009, Jagdhuber2013}.

\subsection{Comparison to prior iterature}

The achieved accuracy levels are encouraging in the context of recent L-band field-scale retrieval studies, but the comparison should be made with care because the target environments differ substantially. Studies such as \cite{Shi2021,Huang2021Agronomy,Koyama2015,Zhu2019Drydown} mainly addressed agricultural scenes with more homogeneous cover conditions and different validation approaches. Most prior L-band PolSAR soil-moisture studies have focused on agricultural fields, where crop type, field boundaries, management practices, and soil conditions are generally better constrained than in engineered mine environments.

Reported accuracies in comparable L-band SAR and PolSAR studies commonly fall in the range of approximately 5--10 volumetric percentage points RMSE. For example, Huang et al. \cite{Huang2021Agronomy} evaluated PALSAR-2 polarimetric decomposition and machine learning retrievals over agricultural fields and reported RMSE values of 9.34 vol.\% for the model-based decomposition approach and 7.70 vol.\% for the Random forest model using SAR and optical imagery. Shi et al. \cite{Shi2021} used L-band multi-incidence and multitemporal PolSAR observations with polarimetric decomposition techniques and reported retrieval errors below 6 vol.\% under agricultural conditions. Zhu et al. \cite{Zhu2019Drydown} reported RMSE values of 0.070~m\(^3\)/m\(^3\) at 25~m pixel scale and 0.056~m\(^3\)/m\(^3\) at paddock scale using time-series multi-angular L-band radar data with a dry-down constraint. Earlier dual-pol ALOS PALSAR and ALOS-2 PALSAR-2 studies have also reported soil-moisture retrieval accuracies on the order of several volumetric percentage points under more controlled agricultural conditions \cite{Koyama2015}.

In this context, the best accuracies obtained in our study, approximately 5.65--5.75 vol.\% RMSE depending on the retrieval family appear competitive.  At the same time, direct numerical comparison with prior studies should not be overemphasized. Some studies used airborne or multi-angular observations, larger numbers of acquisition geometries, explicit roughness or vegetation measurements, optical predictors, or spatial aggregation to field scale. In contrast, the present study relies on a limited nine-date PALSAR-2 time series, sparse point-scale IoT reference measurements, and five contrasting sediment classes within a compact engineered landscape. Our main contribution is therefore not only the absolute RMSE level, but the demonstration that sediment-specific calibration and physically interpretable time-series PolSAR descriptors can produce competitive SSM retrieval performance under complex mine-site conditions.

The comparison also highlights why explicit separation of vegetation and volume-scattering contributions remains difficult. Previous L-band PolSAR studies over vegetation have often relied on multi-angular observations or polarimetric decomposition frameworks to constrain surface, double-bounce, and volume components. In the present single-geometry repeat-pass setting, the available information is more limited, and simple volume-scattering subtraction did not improve the retrieval.

Finally, while comparison to Sentinel-1 based studies in general is beyond the scope of this study, it's noteworthy that for the same site, L-band retrievals provide accuracy of several \% units better than Sentinel-1, as described in the companion paper \cite{hamedianfar2026highresolutionsedimentspecificsurface}, particularly over vegetation covered organic cover. This difference is consistent with the expected stronger soil moisture sensitivity of L-band PolSAR observations under rough or partially vegetated surface conditions, and serves as a site-specific evidence that repeat-pass L-band SAR can provide complementary and, in this case, improved SSM retrieval capability compared to C-band Sentinel-1 imagery.

\subsection{Limitations of the present study}

Our study demonstrates promising high-resolution SM retrieval performance using quad-pol L-band SAR time series, but several limitations should be noted. First, the number of matched PALSAR-2 observations was limited, with the final feature bank containing 90 polarimetric observations distributed across five sediment classes. This is sufficient for a controlled benchmark of compact and interpretable methods, but it limits the reliable training of more complex ML models and makes sediment-specific calibration sensitive to the number of observations per class.

Second, the time series contained only nine repeat-pass PALSAR-2 acquisitions. Although the constant imaging geometry is advantageous for temporal normalization, the number of dates limits the ability to observe the full drying and wetting cycles. The available acquisitions do not  fully sample the driest and wettest possible surface states for each sediment class. This limitation is important for calculating time-series based SMI (dry--wet projection approach), which rely on representative dry and wet state reference PolSAR observations.

Third, the reference SSM observations were obtained from capacitance IoT sensors which measure the soils moisture locally,  while the SAR observations represent spatially averaged scattering signatures. Local variations in sediment texture, compaction, vegetation cover, drainage, and surface roughness may cause the radar footprint to differ from the immediate sensor environment.

Fourth, the study did not include direct field measurements of surface roughness, crusting, cracking, vegetation biomass, or near-surface density at each SAR acquisition date. In general, these variables are known to influence polarimetric signature. Their absence limits the ability to fully separate dielectric moisture effects from structural scattering effects.

Finally, the analysis was performed at a single study site. The site is unique as it contains several contrasting sediment and vegetation cover types, enabling a dedicated study. However, additional validation over other engineered environments would be needed to assess the general applicability of the observed sediment-specific behavior.

\subsection{Implications and outlook} 

Reported  results have several implications for SAR-based SSM retrieval practices. They indicate that L-band PolSAR can provide useful moisture-sensitive information over mineral and organic sediments with absent or moderate vegetation cover, but that the retrieval should be conditioned on these factors. This could be achieved using field and potentially remote-sensing based mapping of soil types and land-cover classes. 
Furthermore, the comparable performance of compact polarimetric representations and  full polarimetric descriptors suggests that L-band SAR missions with systematic repeat-pass observations may support practical SSM retrieval even when full polarimetry is not always available. This is relevant for NISAR and planned ROSE-L, where broad-area time-series observations may provide stronger temporal constraints than the limited tasking-based ALOS-2 PALSAR-2 image stack available in this study. 

Overall, our findings support the development of hybrid retrieval frameworks that combine physically interpretable SAR indices, sediment-specific calibration, and auxiliary environmental predictors. Potential auxiliary data include optical vegetation indices, topographic wetness metrics, drainage information, weather data, freeze--thaw state, and coarse-resolution soil moisture products from radiometer or reanalysis sources. 
Such information may help distinguish true moisture variation from changes in vegetation, roughness, or surface structure. 

Finally, future field campaigns should aim to collect denser spatial reference measurements and additional surface-state information. In particular, surface roughness, cracking, crusting, vegetation cover, and sediment physical properties should be documented together with soil moisture. This would make it possible to test whether the class-specific retrieval functions can be replaced or supplemented by more physically explicit variables, improving both accuracy and transferability.

\section{Conclusion}

This manuscript presents a modelling based framework for L-band PolSAR SSM retrieval relevant to mine sites. 

We explored the performance of L-band quad-pol SAR approaches for SSM retrieval using nine repeat-pass ALOS-2 PALSAR-2 acquisitions and timely in situ IoT soil moisture measurements. The study compared three groups of methods: compact semi-empirical and physics-motivated approaches, time series SMI (dry--wet projection) based retrievals formulated in different polarimetric descriptor spaces, and conventional ML baseline approaches. All methods were evaluated using a common five-fold out-of-fold validation protocol.

Our results show that compact, physically interpretable retrieval approaches remain highly competitive in this complex environment. The best semi-empirical configuration, a cover-specific HH-based soil moisture index augmented with a GVSM-based vegetation descriptor, achieved \(R^2 = 0.67\) and RMSE \(= 5.65\) vol. \%. PolSAR time series based SMI (dry--wet projection) methods achieved comparable performance, with the strongest \([T3]\) diagonal dB and HH/HV dB projection variants reaching approximately \(R^2 = 0.66\) and RMSE \(= 5.67\) vol. \% under sediment-specific calibration. Examined ML methods approached these accuracies, but did not  outperform them, with best temporal and sediment-specific ridge-regression approach achieving \(R^2 = 0.655\) and RMSE \(= 5.75\) vol. \%.

The key observation is that sediment information is essential for reliable SSM retrieval in the investigated setting. Global model fitting performed considerably poorly than the sediment-specific approach across all studied SSM retrieval techniques. This indicates that the relationship between L-band PolSAR observables and volumetric SSM is strongly affected by sediment texture, cover type, roughness, and relevant vegetation structure conditions. The retrieval problem should therefore be treated as a set of sediment- or cover-specific inverse problems rather than as a single universal all-site-wide calibration.

The dry--wet projection experiments further showed that the choice of polarimetric representation is important. Decibel-scaled amplitude and compact descriptor spaces outperformed raw linear and trace-normalized matrix formulations, indicating that the dominant moisture-sensitive trajectory in our repeat-pass L-band dataset is expressed mainly through logarithmic power dynamics. At the same time, adding increasingly rich full-polarimetric descriptors did not improve the retrieval accuracy, suggesting that polarimetric representation with only a few descriptors may retain much of the practically useful information when combined with the sediment-specific calibration.

Vegetation-sensitive descriptors were useful but should be interpreted carefully. The GVSM-inspired radar vegetation descriptor improved the best cover-specific HH-SMI formulation, whereas direct decontamination of HH power by subtracting a scaled HV contribution did not provide a clear benefit. This suggests that cross-polarized component carries not only vegetation scattering component but includes information on sediment structure and moisture-dependent depolarization.

Overall, the study demonstrates that simplified, physics based L-band PolSAR retrievals can provide useful SSM estimates in heterogeneous engineered environments, with accuracy comparable to generic ML approaches. Future work should extend the evaluation to spatially more dense field networks and a wider range of mine sites, longer L-band time series, and auxiliary measurements of roughness, vegetation, cracking, and surface state. Such developments would support more transferable SAR-based soil moisture monitoring for complex engineered landscapes and future L-band missions.

\section*{Acknowledgment}
The MultiMiner project is funded by the European Union’s Horizon Europe research and innovations actions programme under Grant Agreement No. 101091374. We would like to thank Kati Laakso (GTK) for organizing the communication for field work arrangements and Hannele Penson (GTK) for assistance with the field work, and Nordkalk Oy staff including Johanna Huitti for hosting the in situ sensor network and helping with the ground surveys.

\IEEEtriggeratref{35}

\bibliographystyle{IEEEtran}
\bibliography{lib}

\begin{thebibliography}{10}
\providecommand{\url}[1]{#1}
\csname url@samestyle\endcsname
\providecommand{\newblock}{\relax}
\providecommand{\bibinfo}[2]{#2}
\providecommand{\BIBentrySTDinterwordspacing}{\spaceskip=0pt\relax}
\providecommand{\BIBentryALTinterwordstretchfactor}{4}
\providecommand{\BIBentryALTinterwordspacing}{\spaceskip=\fontdimen2\font plus
\BIBentryALTinterwordstretchfactor\fontdimen3\font minus
  \fontdimen4\font\relax}
\providecommand{\BIBforeignlanguage}[2]{{%
\expandafter\ifx\csname l@#1\endcsname\relax
\typeout{** WARNING: IEEEtran.bst: No hyphenation pattern has been}%
\typeout{** loaded for the language `#1'. Using the pattern for}%
\typeout{** the default language instead.}%
\else
\language=\csname l@#1\endcsname
\fi
#2}}
\providecommand{\BIBdecl}{\relax}
\BIBdecl

\bibitem{Hajnsek2003}
I.~Hajnsek, E.~Pottier, and S.~R. Cloude, ``Inversion of surface parameters
  from polarimetric {SAR},'' \emph{IEEE Transactions on Geoscience and Remote
  Sensing}, vol.~41, no.~4, pp. 727--744, 2003.

\bibitem{Hajnsek2009}
I.~Hajnsek, T.~Jagdhuber, H.~Sch{\"o}n, and K.~P. Papathanassiou, ``Potential
  of estimating soil moisture under vegetation cover by means of {PolSAR},''
  \emph{IEEE Transactions on Geoscience and Remote Sensing}, vol.~47, no.~2,
  pp. 442--454, 2009.

\bibitem{Jagdhuber2013}
T.~Jagdhuber, I.~Hajnsek, A.~Bronstert, and K.~P. Papathanassiou, ``Soil
  moisture estimation under low vegetation cover using a multi-angular
  polarimetric decomposition,'' \emph{IEEE Transactions on Geoscience and
  Remote Sensing}, vol.~51, no.~4, pp. 2201--2215, 2013.

\bibitem{Shi2021}
H.~Shi, L.~Zhao, J.~Yang, J.~M. Lopez-Sanchez, J.~Zhao, W.~Sun, L.~Shi, and
  P.~Li, ``Soil moisture retrieval over agricultural fields from {L}-band
  multi-incidence and multitemporal {PolSAR} observations using polarimetric
  decomposition techniques,'' \emph{Remote Sensing of Environment}, vol. 261,
  p. 112485, 2021.

\bibitem{Ulaby1978}
F.~T. Ulaby, P.~P. Batlivala, and M.~C. Dobson, ``Microwave backscatter
  dependence on surface roughness, soil moisture, and soil texture: Part
  i---bare soil,'' \emph{IEEE Transactions on Geoscience Electronics}, vol.~16,
  no.~4, pp. 286--295, 1978.

\bibitem{Oh1992}
Y.~Oh, K.~Sarabandi, and F.~T. Ulaby, ``An empirical model and an inversion
  technique for radar scattering from bare soil surfaces,'' \emph{IEEE
  Transactions on Geoscience and Remote Sensing}, vol.~30, no.~2, pp. 370--381,
  1992.

\bibitem{Ulaby1996}
F.~T. Ulaby, P.~C. Dubois, and J.~van Zyl, ``Radar mapping of surface soil
  moisture,'' \emph{Journal of Hydrology}, vol. 184, no. 1--2, pp. 57--84,
  1996.

\bibitem{Dubois1995}
P.~C. Dubois, J.~van Zyl, and T.~Engman, ``Measuring soil moisture with imaging
  radars,'' \emph{IEEE Transactions on Geoscience and Remote Sensing}, vol.~33,
  no.~4, pp. 915--926, 1995.

\bibitem{Shi1997}
J.~Shi, J.~Wang, A.~Y. Hsu, P.~E. O'Neill, and E.~T. Engman, ``Estimation of
  bare surface soil moisture and surface roughness parameter using l-band sar
  image data,'' \emph{IEEE Transactions on Geoscience and Remote Sensing},
  vol.~35, no.~5, pp. 1254--1266, 1997.

\bibitem{Attema1978}
E.~P.~W. Attema and F.~T. Ulaby, ``Vegetation modeled as a water cloud,''
  \emph{Radio Science}, vol.~13, no.~2, pp. 357--364, 1978.

\bibitem{ElHajj2016}
M.~El~Hajj, N.~Baghdadi, M.~Zribi, G.~Belaud, B.~Cheviron, D.~Courault, and
  F.~Charron, ``Soil moisture retrieval over irrigated grassland using x-band
  sar data,'' \emph{Remote Sensing of Environment}, vol. 176, pp. 202--218,
  2016.

\bibitem{Singh2023}
S.~K. Singh, R.~Prasad, P.~K. Srivastava, S.~A. Yadav, V.~P. Yadav, and
  J.~Sharma, ``Incorporation of first-order backscattered power in water cloud
  model for improving the leaf area index and soil moisture retrieval using
  dual-polarized sentinel-1 sar data,'' \emph{Remote Sensing of Environment},
  vol. 296, p. 113756, 2023.

\bibitem{Yadav2020}
V.~P. Yadav, R.~Prasad, R.~Bala, and A.~K. Vishwakarma, ``An improved inversion
  algorithm for spatio-temporal retrieval of soil moisture through modified
  water cloud model using c-band sentinel-1a sar data,'' \emph{Computers and
  Electronics in Agriculture}, vol. 173, p. 105447, 2020.

\bibitem{Freeman1998}
A.~Freeman and S.~L. Durden, ``A three-component scattering model for
  polarimetric {SAR} data,'' \emph{IEEE Transactions on Geoscience and Remote
  Sensing}, vol.~36, no.~3, pp. 963--973, 1998.

\bibitem{Yamaguchi2005}
Y.~Yamaguchi, T.~Moriyama, M.~Ishido, and H.~Yamada, ``Four-component
  scattering model for polarimetric {SAR} image decomposition,'' \emph{IEEE
  Transactions on Geoscience and Remote Sensing}, vol.~43, no.~8, pp.
  1699--1706, 2005.

\bibitem{vanZyl2011}
J.~J. van Zyl, M.~Arii, and Y.~Kim, ``Model-based decomposition of polarimetric
  {SAR} covariance matrices constrained for nonnegative eigenvalues,''
  \emph{IEEE Transactions on Geoscience and Remote Sensing}, vol.~49, no.~9,
  pp. 3452--3459, 2011.

\bibitem{Wagner1999}
W.~Wagner, G.~Lemoine, and H.~Rott, ``A method for estimating soil moisture
  from ers scatterometer and soil data,'' \emph{Remote Sensing of Environment},
  vol.~70, no.~2, pp. 191--207, 1999.

\bibitem{Balenzano2011}
A.~Balenzano, F.~Mattia, G.~Satalino, and M.~W.~J. Davidson, ``Dense temporal
  series of c- and l-band sar data for soil moisture retrieval over
  agricultural crops,'' \emph{IEEE Journal of Selected Topics in Applied Earth
  Observations and Remote Sensing}, vol.~4, no.~2, pp. 439--450, 2011.

\bibitem{Balenzano2021}
A.~Balenzano, F.~Mattia, G.~Satalino, F.~P. Lovergine, D.~Palmisano, J.~Peng,
  P.~Marzahn, U.~Wegm{\"u}ller, O.~Cartus, K.~Dabrowska-Zieli{\'n}ska, J.~P.
  Musial, M.~W.~J. Davidson, V.~R.~N. Pauwels, M.~H. Cosh, H.~McNairn, J.~T.
  Johnson, J.~P. Walker, S.~H. Yueh, D.~Entekhabi, Y.~H. Kerr, and T.~J.
  Jackson, ``Sentinel-1 soil moisture at 1 km resolution: A validation study,''
  \emph{Remote Sensing of Environment}, vol. 263, p. 112554, 2021.

\bibitem{Zhu2022}
L.~Zhu, R.~Si, X.~Shen, and J.~P. Walker, ``An advanced change detection method
  for time-series soil moisture retrieval from sentinel-1,'' \emph{Remote
  Sensing of Environment}, vol. 279, p. 113137, 2022.

\bibitem{He2016}
L.~He, R.~Panciera, M.~A. Tanase, J.~P. Walker, and Q.~Qin, ``Soil moisture
  retrieval in agricultural fields using adaptive model-based polarimetric
  decomposition of sar data,'' \emph{IEEE Transactions on Geoscience and Remote
  Sensing}, vol.~54, no.~8, pp. 4445--4460, 2016.

\bibitem{Wang2017}
H.~Wang, R.~Magagi, and K.~Goita, ``Comparison of different polarimetric
  decompositions for soil moisture retrieval over vegetation covered
  agricultural area,'' \emph{Remote Sensing of Environment}, vol. 199, pp.
  120--136, 2017.

\bibitem{Jagdhuber2016}
T.~Jagdhuber, ``An approach to extended {Fresnel} scattering for modeling of
  depolarizing soil--trunk double-bounce scattering,'' \emph{Remote Sensing},
  vol.~8, no.~10, p. 818, 2016.

\bibitem{antropov2011}
O.~Antropov, Y.~Rauste, and T.~H{\"a}me, ``Volume scattering modeling in
  {PolSAR} decompositions: Study of {ALOS} {PALSAR} data over boreal forest,''
  \emph{IEEE Transactions on Geoscience and Remote Sensing}, vol.~49, no.~10,
  pp. 3838--3848, 2011.

\bibitem{Abowarda2021}
A.~S. Abowarda, L.~Bai, C.~Zhang, D.~Long, X.~Li, Q.~Huang, and Z.~Sun,
  ``Generating surface soil moisture at 30 m spatial resolution using both data
  fusion and machine learning toward better water resources management at the
  field scale,'' \emph{Remote Sensing of Environment}, vol. 255, p. 112301,
  2021.

\bibitem{Huang2021Agronomy}
X.~Huang, B.~Ziniti, M.~H. Cosh, M.~Reba, J.~Wang, and N.~Torbick,
  ``Field-scale soil moisture retrieval using {PALSAR-2} polarimetric
  decomposition and machine learning,'' \emph{Agronomy}, vol.~11, no.~1, p.~35,
  2021.

\bibitem{Bhogapurapu2022}
N.~Bhogapurapu, S.~Dey, D.~Mandal, A.~Bhattacharya, L.~Karthikeyan, H.~McNairn,
  and Y.~S. Rao, ``Soil moisture retrieval over croplands using dual-pol l-band
  grd sar data,'' \emph{Remote Sensing of Environment}, vol. 271, p. 112900,
  2022.

\bibitem{Fan2025}
D.~Fan, T.~Zhao, X.~Jiang, A.~Garc{\'i}a-Garc{\'i}a, T.~Schmidt, L.~Samaniego,
  S.~Attinger, H.~Wu, Y.~Jiang, J.~Shi, L.~Fan, B.~H. Tang, W.~Wagner,
  W.~Dorigo, A.~Gruber, F.~Mattia, A.~Balenzano, L.~Brocca, T.~Jagdhuber, J.-P.
  Wigneron, C.~Montzka, and J.~Peng, ``A sentinel-1 sar-based global 1-km
  resolution soil moisture data product: Algorithm and preliminary
  assessment,'' \emph{Remote Sensing of Environment}, vol. 318, p. 114579,
  2025.

\bibitem{Karthikeyan2017}
L.~Karthikeyan, M.~Pan, N.~Wanders, D.~N. Kumar, and E.~F. Wood, ``Four decades
  of microwave satellite soil moisture observations: Part 2. product validation
  and inter-satellite comparisons,'' \emph{Advances in Water Resources}, vol.
  109, pp. 236--252, 2017.

\bibitem{Kornelsen2013}
K.~C. Kornelsen and P.~Coulibaly, ``Advances in soil moisture retrieval from
  synthetic aperture radar and hydrological applications,'' \emph{Journal of
  Hydrology}, vol. 476, pp. 460--489, 2013.

\bibitem{Lamichhane2025}
M.~Lamichhane, S.~Mehan, and K.~R. Mankin, ``Soil moisture prediction using
  remote sensing and machine learning algorithms: A review on progress,
  challenges, and opportunities,'' \emph{Remote Sensing}, vol.~17, no.~14, p.
  2397, 2025.

\bibitem{Zhu2025TGRS}
L.~Zhu, J.~Dai, J.~Jin, S.~Yuan, Z.~Xiong, and J.~P. Walker, ``Are the current
  expectations for sar remote sensing of soil moisture using machine learning
  overoptimistic?'' \emph{IEEE Transactions on Geoscience and Remote Sensing},
  vol.~63, pp. 1--15, 2025.

\bibitem{antropovigarss2024}
O.~Antropov, M.~Molinier, L.~Seitsonen, A.~Hamedianfar, M.~Middleton,
  K.~Laakso, H.~Sutinen, and P.~Liwata-Kenttälä, ``Continuous ground moisture
  monitoring at limestone quarry using multi-sensor sar images and in situ iot
  sensors,'' in \emph{IGARSS 2024 - 2024 IEEE International Geoscience and
  Remote Sensing Symposium}, 2024, pp. 3392--3395.

\bibitem{hamedianfar2026highresolutionsedimentspecificsurface}
\BIBentryALTinterwordspacing
A.~Hamedianfar, O.~Antropov, M.~Molinier, U.~Salmela, H.~Kukkula, L.~Seitsonen,
  P.~Liwata-Kenttälä, and M.~Middleton, ``High resolution sediment-specific
  surface soil moisture retrieval using sentinel-1 time series and auxiliary
  data,'' 2026, in Review. [Online]. Available:
  \url{https://doi.org/10.48550/arXiv.2606.24364}
\BIBentrySTDinterwordspacing

\bibitem{shimada2009}
M.~Shimada, O.~Isoguchi, T.~Tadono, and K.~Isono, ``Palsar radiometric and
  geometric calibration,'' \emph{IEEE Transactions on Geoscience and Remote
  Sensing}, vol.~47, no.~12, pp. 3915--3932, 2009.

\bibitem{rauste2007}
Y.~Rauste, A.~Lonnqvist, M.~Molinier, J.-B. Henry, and T.~Hame,
  ``Ortho-rectification and terrain correction of polarimetric sar data applied
  in the alos/palsar context,'' in \emph{2007 IEEE International Geoscience and
  Remote Sensing Symposium}, 2007, pp. 1618--1621.

\bibitem{small2011}
D.~Small, ``Flattening gamma: Radiometric terrain correction for sar imagery,''
  \emph{IEEE Transactions on Geoscience and Remote Sensing}, vol.~49, no.~8,
  pp. 3081--3093, 2011.

\bibitem{NLS2025}
{NLS}, ``{Elevation model 2 m},''
  \url{https://www.maanmittauslaitos.fi/en/maps-and-spatial-data/datasets-and-interfaces/product-descriptions/elevation-model-2-m},
  National Land Survey of Finland, 2025, accessed: 2025-09-27.

\bibitem{praks2009}
J.~Praks, E.~C. Koeniguer, and M.~T. Hallikainen, ``Alternatives to target
  entropy and alpha angle in sar polarimetry,'' \emph{IEEE Transactions on
  Geoscience and Remote Sensing}, vol.~47, no.~7, pp. 2262--2274, 2009.

\bibitem{kim2009}
Y.~Kim and J.~J. van Zyl, ``A time-series approach to estimate soil moisture
  using polarimetric radar data,'' \emph{IEEE Transactions on Geoscience and
  Remote Sensing}, vol.~47, no.~8, pp. 2519--2527, 2009.

\bibitem{mandal2020}
D.~Mandal, D.~Ratha, A.~Bhattacharya, V.~Kumar, H.~McNairn, Y.~S. Rao, and
  A.~C. Frery, ``A radar vegetation index for crop monitoring using compact
  polarimetric sar data,'' \emph{IEEE Transactions on Geoscience and Remote
  Sensing}, vol.~58, no.~9, pp. 6321--6335, 2020.

\bibitem{hoerl1970ridge}
A.~E. Hoerl and R.~W. Kennard, ``Ridge regression: Biased estimation for
  nonorthogonal problems,'' \emph{Technometrics}, vol.~12, no.~1, pp. 55--67,
  1970.

\bibitem{cover1967nearest}
T.~Cover and P.~Hart, ``Nearest neighbor pattern classification,'' \emph{IEEE
  Transactions on Information Theory}, vol.~13, no.~1, pp. 21--27, 1967.

\bibitem{tomppo2008}
E.~Tomppo, H.~Olsson, G.~Ståhl, M.~Nilsson, O.~Hagner, and M.~Katila,
  ``Combining national forest inventory field plots and remote sensing data for
  forest databases,'' \emph{Remote Sensing of Environment}, vol. 112, no.~5,
  pp. 1982--1999, 2008.

\bibitem{antropov2017}
\BIBentryALTinterwordspacing
O.~Antropov, Y.~Rauste, T.~Häme, and J.~Praks, ``Polarimetric {ALOS PALSAR}
  time series in mapping biomass of boreal forests,'' \emph{Remote Sensing},
  vol.~9, no.~10, 2017. [Online]. Available:
  \url{https://www.mdpi.com/2072-4292/9/10/999}
\BIBentrySTDinterwordspacing

\bibitem{khatami2016}
\BIBentryALTinterwordspacing
R.~Khatami, G.~Mountrakis, and S.~V. Stehman, ``A meta-analysis of remote
  sensing research on supervised pixel-based land-cover image classification
  processes: General guidelines for practitioners and future research,''
  \emph{Remote Sensing of Environment}, vol. 177, pp. 89--100, 2016. [Online].
  Available:
  \url{https://www.sciencedirect.com/science/article/pii/S0034425716300578}
\BIBentrySTDinterwordspacing

\bibitem{smola2004tutorial}
A.~J. Smola and B.~Sch{\"o}lkopf, ``A tutorial on support vector regression,''
  \emph{Statistics and Computing}, vol.~14, pp. 199--222, 2004.

\bibitem{breiman2001random}
L.~Breiman, ``Random forests,'' \emph{Machine Learning}, vol.~45, pp. 5--32,
  2001.

\bibitem{chen2016xgboost}
T.~Chen and C.~Guestrin, ``Xgboost: A scalable tree boosting system,'' in
  \emph{Proceedings of the 22nd ACM SIGKDD International Conference on
  Knowledge Discovery and Data Mining}, 2016, pp. 785--794.

\bibitem{Koyama2015}
C.~N. Koyama, K.~Schneider, and M.~Sato, ``Development of a biomass corrected
  soil moisture retrieval model for dual-polarization {ALOS-2} data based on
  {ALOS}/{PALSAR} and {Pi-SAR-L2} observations,'' in \emph{Proceedings of the
  IEEE International Geoscience and Remote Sensing Symposium (IGARSS)}, 2015,
  pp. 1316--1319.

\bibitem{Zhu2019Drydown}
L.~Zhu, J.~P. Walker, L.~Tsang, H.~Huang, N.~Ye, and C.~R{\"u}diger, ``Soil
  moisture retrieval from time series multi-angular radar data using a dry down
  constraint,'' \emph{Remote Sensing of Environment}, vol. 231, p. 111237,
  2019.

\end{thebibliography}

\end{document}